uses epsf.sty, epsfig.sty and epsf.tex
\documentstyle[11pt,epsfig,bezier]{article}      

  \newcommand{\dreij}[6]{ \left(
                          \matrix{#1 & #2 & #3 \cr     
                                  #4 & #5 & #6 }
                          \right)   }

  \newcommand{\nn}{\nonumber}                          

  \newcommand{\be}{\begin{equation}}                   

  \newcommand{\ee}{\end{equation}}                     

  \newcommand{\benn}{\begin{displaymath}}              

  \newcommand{\eenn}{\end{displaymath}}                

  \newcommand{\bee}{\begin{samepage}                   
                    \begin{eqnarray}}                  

  \newcommand{\eee}{
                    \end{eqnarray}                     
                    \end{samepage}}

  \newcommand{\tr}{\mbox{tr\,}}                          

  \newcommand{\Tr}{\mbox{Tr\,}}                          

  \newcommand{\Det}{\mbox{Det\,}}                        

  \newcommand{\ket}[1]{| #1 \rangle}                   

  \newcommand{\bra}[1]{\langle #1 \, |}                

   \newcommand{\threehalf}{{3 \over 2}}

  \newcommand{\mbf}[1]{{\mbox{\bf #1}}}                
\newcommand{\semiclass}{\ \stackrel{\rm s.c.}{\longrightarrow}\ }
\title{Quantization of Hyperbolic $N$-Sphere Scattering Systems
       in Three Dimensions}

\author{Michael Henseler\thanks{present address: 
        Max-Planck-Institut f\"{u}r Physik
        komplexer Systeme, Bayreuther Str.\,40, 
        Haus~16, 01187 Dresden, Germany,
        {michael@mpipks-dresden.mpg.de}},
        Andreas Wirzba\thanks{
        Andreas.Wirzba@physik.th-darmstadt.de}\\
        {\it Institut f\"ur Kernphysik, TH Darmstadt,
             Schlo\ss gartenstra\ss e 9},\\
        {\it 64289 Darmstadt, Germany},\\
        and\\
        Thomas Guhr\thanks{present address:
        Max-Planck-Institut f\"ur Kernphysik,
        Postfach 103980, 69029 Heidelberg, Germany,
        guhr@mickey.mpi-hd.mpg.de}\\
        {\it  Niels Bohr Institute, Blegdamsvej 17},\\
        {\it  2100 Copenhagen \O, Denmark}
        }

\begin{document}

\date{\today} 

\maketitle

\begin{abstract}
Most discussions of chaotic scattering systems are devoted to
two-dimensional systems. It is of considerable interest to extend
these studies to the, in general, more realistic case of three
dimensions. In this context, it is conceptually important to
investigate the quality of semiclassical methods as a function of the
dimensionality. As a model system, we choose various three dimensional
generalizations of the famous three disk problem which played a
central role in the study of chaotic scattering in two dimensions.  We
present a quantum-mechanical treatment of the hyperbolic scattering of
a point particle off a finite number of non-overlapping and non-touching
hard spheres in three dimensions. We derive expressions for the
scattering matrix {\bf S} and its determinant. The determinant of {\bf
S} decomposes into two parts, the first one contains 
the product of the determinants
of the individual one-sphere {\bf S}-matrices and the second one is
given by a ratio involving the determinants of a characteristic
KKR-type matrix and its conjugate.  We justify our approach by showing
that all formal manipulations in these derivations are correct and
that all the determinants involved which are of infinite dimension
exist.  Moreover, for all complex wave numbers, we conjecture a direct
link between the quantum-mechanical and semiclassical descriptions:
The semiclassical limit of the cumulant expansion of the KKR-type
matrix is given by the Gutzwiller-Voros zeta function plus
diffractional corrections in the curvature expansion.  
This connection is direct since it is not
based on any kind of subtraction scheme involving bounded reference
systems.
We present numerically
computed resonances and compare them with the corresponding data for
the similar two-dimensional $N$-disk systems and with semiclassical
calculations.
\end{abstract}
\newpage



\section{Introduction}

Many if not most of the concepts in quantum chaos were developed and
are being applied to two-dimensional systems. This is due to the
relative simplicity of those systems as compared to three-dimensional
ones. To find the periodic orbits in a generically three-dimensional
geometry, obviously requires much more work than performing the same
type of analysis in a two-dimensional billiard, say. However, since
the real world is three-dimensional, an extension of chaos studies to
more realistic systems is called for. The celebrated Hydrogen Atom in
a strong magnetic field can be reduced to an, effectively, two
dimensional system due to the cylindrical symmetry of the problem.
Similar simplifications also exist in other systems.  Recently, a
full-fledged study of the three-dimensional Sinai-billiard has been
performed~\cite{Sinai_3d_qm}.  Here, we present a study of a chaotic
scattering system in three dimensions. We investigate several
generalizations of the two-dimensional three disk problem.  We chose
three-dimensional versions of this system since it played an important
role in the development of many concepts and methods in chaotic
scattering. In order to describe $N$-sphere scattering systems
quantum-mechanically we extend the methods of Refs.\,\cite{GaRiqm,
unspaper} to three dimensions.  A related approach was already used
in Ref.\,\cite{LlSm} to describe the multiple scattering of an electron in
non-overlapping muffin-tin potentials.  In Ref.\,\cite{JSS} the
scattering of a point particle on non-symmetric configurations of
point-scatterers in three dimensions was investigated.

In general, hyperbolic or even chaotic {\em scattering} systems have
some advantages compared with {\em bounded} systems, if one is
interested in the quantum-mechanical {\em and} the semiclassical
description of a classically chaotic problem.  Chaotic bounded systems
are normally plagued by the presence of non-isolated, non-hyperbolic
bouncing ball orbits (see, e.g., the Sinai-billiard or the stadium
billiard), the importance of very long periodic orbits, and the
problem that without fine-tuning a la Berry and
Keating~\cite{berry_keats} the semiclassical Gutzwiller-Voros zeta
function~\cite{gutzwiller,voros88} would not predict {\em real-valued}
energy eigenvalues. In contrast, the geometry of hyperbolic scattering
systems can easily be chosen such that bouncing ball orbits are
absent. Furthermore, the contributions of longer periodic orbits to
the scattering matrix are automatically suppressed relatively to the
shorter ones and no fine-tuning is necessary in order to predict
scattering resonances as they are anyhow complex-valued.

Two-dimensional $N$-disk and three-dimensional $N$-sphere systems are
examples of such scattering systems which are simple enough to be
studied {\em in all detail} semiclassically and quantum-mechanically.
In the past the two-dimensional Sinai-billiard, which can be
interpreted as the scattering on an infinite, regular array of equal
disks, has been quantized~\cite{Berry_KKR} using the
Korringa-Kohn-Rostoker (KKR) method~\cite{KKR}.  More recently the
scattering of a point particle on hard disks in two dimensions~\cite{Eck} 
has been studied classically~\cite{GaRikl,PoCh},
quantum-mechanically~\cite{GaRiqm, unspaper} and
semiclassically~\cite{GaRiskl,CE89,fredh,Pinball} using the
techniques of periodic orbit theory~\cite{gutzwiller}.  The range of
validity of the purely geometrical semiclassical input has been
investigated in Refs.\,\cite{AW_testof,CVW96}.  In
Refs.\cite{alonso,gasp_hbar,vattay_hbar} $\hbar$-corrections to the
geometrical periodic orbits were constructed, whereas the authors of
Refs.\cite{vwr_prl,vwr_jap,rvw_jsp} 
extended the Gutzwiller-Voros zeta function to
include diffractive creeping periodic orbits as well.  Recently a
formalism which also includes the limit of small disks ($ka \ll 1$,
$k$: wave number, $a$: radius of the disk) was
presented~\cite{RoWhWi}.  In Ref.\,\cite{unspaper} the connection
between the quantum-mechanical and semiclassical descriptions of
$N$-disk scattering systems has been investigated in
detail. 
On the experimental side, the scattering
on two equal disks was investigated using microwave
cavities~\cite{Kudrolli_Sridhar}.  In Ref.\,\cite{Sinai_2d_qm} the
two-dimensional Sinai-billiard was treated in a scattering approach
and in Ref.\,\cite{Sinai_2d_diff} diffractive effects were considered.

Here we focus on the analogous scattering systems of $N$ spheres in
three dimensions.  These systems are still simple enough to be treated
quantum-mechanically, semiclassically and classically.  The
quantum-mechanical description of $N$-sphere systems is similar to the
two-dimensional $N$-disk case, as essentially the same approach can be
used. Also the semiclassical description with purely geometrical input
is similar to the two-dimensional case. But diffractional corrections
should be substantially different as the corresponding creeping orbits
are now extrema on two-dimensional manifolds instead of
one-dimensional ones. A detailed quantum-mechanical description of
$N$-sphere scattering systems is presented and the connection with the
semiclassical treatment is investigated.

The paper is organized as follows. In Sec.\,2, we derive an explicit
expression for the scattering matrix within the framework of
stationary scattering theory using Green's theorem. We do this in some
detail to make the paper self-contained. In Sec.\,3, the determinant
of the scattering matrix is re-written as a product of an
incoherent part and a coherent part. We show that the scattering
resonances are given by the zeros of a KKR-type matrix.  Moreover, we
conjecture a direct link between the quantum mechanical and
semiclassical descriptions of $N$-sphere systems involving the
Gutzwiller-Voros zeta function plus diffractional corrections. This
connection is valid for all complex wave numbers.  In Sec.\,4, we give
a proof that all formal manipulations performed in the preceding
sections are allowed and that the determinants of the infinite
matrices involved are well defined. In Sec.\,5, we present numerical
results on \mbox{two-,} three- and four-sphere scattering systems and compare
them with the analogous two- and three-disk systems.  In Sec.\,6, we
summarize our results and give an outlook.


\section{Calculation of the Scattering Matrix}

We describe the scattering of a point particle on $N$ hard spheres
within the framework of stationary scattering theory following the
methods of Berry~\cite{Berry_KKR} and Gaspard and Rice~\cite{GaRiqm}.
In Sec.\,2.1, we define the scattering matrix and outline our
approach. In Sec.\,2.2, the elements of the scattering matrix are
worked out explicitly.


\subsection{Definitions and General Concepts}

To describe a generic configuration of $N$ spheres we use the
following notation: $j \in \{1,\ldots,N\}$ specifies one spherical
scatterer with radius $a_j$. $R_{jj'}$ denotes the distance between
the centers of the spheres $j$ and $j'$. To specify positions we use
$N+1$ coordinate systems: First we choose a global coordinate system
$(x,y,z)$ whose origin is situated at an arbitrary point in the
neighbourhood of the $N$ scatterers.  This point is chosen to be the
center of the large sphere of the integration volume used in Green's
theorem (see Sec.\,2.2). If the scattering configuration possesses
symmetries the origin of the global system is placed at the symmetry
center of the entire system. In order to perform symmetry reductions
we introduce $N$ local coordinate systems,
$(x^{(j)},y^{(j)},z^{(j)})$, whose origins lie at the centers of the
$N$ spheres. The axes of these coordinate systems are chosen in such a
way that the symmetry of the entire configuration is respected.  The
vector from the origin of the global coordinate system to the center
of the $j$-th sphere is called $\vec{s}_j$ and it is measured in the
global system. All the vectors in the local coordinate system of sphere $j$ 
are measured relative to this vector.
$\hat{R}_{jj'}^{(j)} \equiv \vec{R}_{jj'}^{(j)} /
R_{jj'}$ denotes the unit vector from the center of the sphere $j$ to
that of $j'$ and it is measured in the $(j)$-system. In general,
vectors with an upper index $(j)$ are measured in the $(j)$-system,
vectors without upper index are measured in the global system.

A solution of the time-independent Schr\"odinger equation fulfills
\bee \label{Schroedinger-1}
 \left( \vec{\nabla}^2 + \vec{k}{}^2 \right) \psi(\vec{r}\,) & = & 0 \; ,
       \quad \mbox{$\vec{r}$ outside the $N$ spheres,}   \\
 \psi(\vec{r}\,) & = & 0 \; , \quad \mbox{$\vec{r}$ on the surfaces 
of the spheres.}
        \nn
\eee     
The energy of the particle is $\hbar^2 \vec{k}{}^2 / 2m$ and $\vec{k}$ is the
wave vector of the incident wave. 
We expand the wave function $\psi(\vec{r}\,)$ in a basis of
eigenfunctions of angular momentum
\be  \label{psi_zerl}       {\psi}(\vec{r}\,) =
  \sum_{l = 0}^\infty \sum_{m = -l}^l \psi_{l m}^k(\vec{r}\,)
                  Y_{lm}^{\ast}(\hat{k}) 
\ee
where $k$ and $\hat{k}$ are the length and the solid angle of
the wave vector, respectively.
Because of this expansion, we construct solutions of the
Schr\"odinger equation for the basis functions 
\be  \label{psi_lm_dgl}
     (\vec{\nabla}{}^2 + {\vec{k}}^2) \psi_{l m}^k(\vec{r}\,) = 0 \, .
\ee
Asymptotically for large distances from the scatterers ($kr \to
\infty$) the spherical components $\psi_{l m}^k$ can be written as a
superposition of in-coming and out-going spherical waves, 
\bee
 \label{Def_S} 
 \psi_{l m}^k(\vec{r}\,)
 & \sim & \frac{2 \pi}{kr} \sum_{l'=0}^\infty \sum_{m'=-l'}^{l'} \{
 i^{l'} [ {e^{-i(kr - \frac{l+1}{2}\pi)} \delta_{ll'}\delta_{mm'}} \\ &
 & \qquad\mbox{}+\mbf{S}_{lm,l'm'} e^{i(kr - \frac{l+1}{2} \pi)} ]
 Y_{l'm'}(\hat{r}) \} \; .  \nn 
\eee 
This equation defines the scattering matrix $\mbf{S}$.
Its elements $\mbf{S}_{lm,l'm'}$ describes the scattering of an
in-coming wave with angular momentum $l,m$ into an out-going wave with
angular momentum $l',m'$.
If there are no scatterers, we have $\mbf{S} = \mbf{1}$ and the asymptotic
expression of a plane wave $e^{i \vec{k} \cdot \vec{r}}$ is
recovered. As is well known, the scattering matrix {\bf S}
is unitary because of probability conservation.

To derive an explicit expression for the {\bf S}-matrix,
 we use the free Green's function, 
\be     \label{Green_fkt}
    (\vec{\nabla}{}^2+{\vec{k}}^2)G(\vec{r},\vec{r}{\,}')  =  -4\pi
                      {\delta}^3(\vec{r}-\vec{r}{\,}')    \quad , \quad
                 G(\vec{r},\vec{r}{\,}')=G(\vec{r}{\,}',\vec{r}\,)
\ee
\be
     G(\vec{r},\vec{r}{\,}')  = 
   4 \pi ik \sum_{l=0}^\infty \sum_{m=-l}^l j_l(kr_<) \, h_l^{(1)}(kr_>)
           \,  Y_{lm}^\ast (\hat{r}) Y_{lm}(\hat{r}') \; ,
\ee
where $r_<$ ($r_>$) denotes the magnitude of the shorter (longer) of the two
vectors $\vec{r}$ and $\vec{r}{\,}'$ whose directions can be expressed in any
coordinate system. 
The Green's function and the components $\psi_{l m}^k(\vec{r})$ of the 
wave function are inserted in Green's theorem which yields
\bee
    \int_V d^3r \, \left[
\psi_{l m}^k(\vec{r}\,) \,(\vec{\nabla}{}^2+{\vec{k}}^2)
    G(\vec{r},\vec{r}{\,}') - G(\vec{r},\vec{r}{\,}')\,
(\vec{\nabla}{}^2+{\vec{k}}^2)      \psi_{l m}^k(\vec{r}\,)\right] \nn \\
=     -4 \pi \int_V d^3r \, \psi_{l m}^k(\vec{r}\,)
{\delta}^3(\vec{r}-\vec{r}{\,}')        \nn   \\
    = \int_{\partial V} \, d\vec{a} \cdot 
  \left[\psi_{l m}^k(\vec{a}) \vec{\nabla}
    G(\vec{a},\vec{r}{\,}')
       - G(\vec{a},\vec{r}{\,}') \vec{\nabla} \psi_{l m}^k(\vec{a})
\right] \nn \\
                      \label{Green}
   =  \cases{ 0 & $\vec{r}{\,}' \notin V$, \cr
             -4 \pi \psi_{l m}^k(\vec{r}{\,}') & $\vec{r}{\,}' \in V$, }
\eee
where $V$ denotes the integration volume and $\partial V$ is its
boundary.  The appropriate volume of integration $V$ consists of a
large sphere centered at an arbitrary point in the neighbourhood of the
$N$ scatterers.
Its radius is chosen to be so large that asymptotic formulae like
Eq.\,(\ref{Def_S}) hold for points far away from the origin but still
inside the integration volume.
{}From this large sphere we exclude $N$ spheres whose centers coincide
with those of the $N$ scatterers and whose radii are larger by an
infinitesimal amount $\epsilon$ than the corresponding radii of the
scatterers. 

\subsection{Computation of the Matrix Elements}

Equation~(\ref{Green}), understood in the limit
$\epsilon \to 0$,  can now be applied to two different
cases as we can choose the point $\vec{r}{\,}'$ to be outside or
inside the volume $V$.  In the first case $\vec{r}{\,}'$ is chosen to
be some point on the surface of the scatterer $j$ and therefore
outside the integration volume $V$.  In the second case $\vec{r}{\,}'$
is chosen to lie inside the volume $V$ and its modulus $r'$ is taken
to be so large that asymptotic formulae like Eq.\,(\ref{Def_S})
hold. The boundary $\partial V$ decomposes into $N\mbox{+}1$ disjoint
parts: The outer boundary of the large sphere, $\partial_\infty V$, and
the $N$ surfaces $\partial_j V$ of the excluded spheres, which
coincide with the scatterers in the limit $\epsilon \to 0$.

{\bf First case:} $\vec{r}{\,}' \equiv \vec{r}_j \in \mbox{boundary 
of the $j^{\rm \, th}$
scatterer}$ \\ \nopagebreak
All components ${\psi}^{k}_{lm}$ vanish on the surfaces of the
scatterers, but the gradient of the wave function on these surfaces is
nonzero.  Its normal component can be expanded in spherical harmonics,
$ Y_{lm}(\hat{r}^{(j)}_j)$,
defined on the surface of the $j^{\rm\, th}$ scatterer,
\be    \label{RB}    
 \vec{n}_j \cdot \vec{\nabla} \psi^k_{lm}(\vec{r}_j)  \equiv 
     \sum_{l'=0}^\infty \sum_{m'=-l'}^{l'} \mbf{A}^j_{lm,l'm'}
             Y_{l'm'}(\hat{r}^{(j)}_j) \; 
\ee
where $\vec{n}_j$ denotes a unit vector perpendicular to $\partial V$ pointing
outside $V$. The orientation $\hat{r}^{(j)}_{j}$ is measured in the
$(j)$-system. The unknown coefficients $\mbf{A}_{lm,l'm'}^j$ 
are then uniquely determined through Green's theorem (\ref{Green}).
In this way the gradient of the wave function is characterized by the
matrix $\mbf{A}^j$ which depends on the sphere label $j$ and whose 
matrix elements are specified by angular momentum
quantum numbers. 
Collecting everything, we arrive at a compact matrix formulation which
expresses the gradient matrix $\bf{A}$ in terms of two matrices $\bf{C}$
and $\bf{M}$
\be   \label{CAM}
   \mbf{C}^{j} = \mbf{A}^{j'} \cdot \mbf{M}^{j'j} \; .
\ee
where we use the Einstein-summation convention for the sphere 
indices $j$ and $j'$. 
The matrix elements of $\mbf{C}^j$ and $\mbf{M}^{jj'}$ 
read 
\bee  \label{C_N_ball_new}
   \mbf{C}^j_{lm,l'm'} & = & \frac{(4\pi)^{\frac{3}{2}}}{ika^2_j} 
     \sum_{l_1=0}^\infty
     \sum_{\tilde{m}=-l'}^{l'} (-1)^m\, i^{l_1+l'}  \\ & &
   \quad \mbox{} \times \sqrt{(2l+1)(2l_1+1)(2l'+1)}\dreij{l_1}{l'}{l}{0}{0}{0}
        \dreij{l_1}{l'}{l}{m-\tilde{m}}{\tilde{m}}{-m}    \nn \\ & &
   \quad \mbox{} \times 
    \frac{j_{l_1}(ks_j)}{h^{(1)}_{l'}(ka_j)} Y_{l_1,m-\tilde{m}}(\hat{s}_j)
         D^{l'}_{m'\tilde{m}}(gl,j)  \;    \nn
\eee
and
\bee       \label{M_N_ball_new}
   \mbf{M}^{jj'}_{lm,l'm'} & = & \delta^{jj'} \delta_{ll'} \delta_{mm'}
             \quad  \\ & &
            \quad \mbox{}+(1-\delta^{jj'})  
   \left( \frac{a_j}{a_{j'}} \right)^2
               \sqrt{4\pi} (-1)^{m}
      \sum_{l_1=0}^\infty \sum_{\tilde{m}=-l'}^{l'} i^{l_1+l'-l} \nn \\ & &
   \qquad \mbox{} \times 
 \sqrt{(2l+1)(2l_1+1)(2l'+1)}\dreij{l_1}{l'}{l}{0}{0}{0}
        \dreij{l_1}{l'}{l}{m-\tilde{m}}{\tilde{m}}{-m}    \nn \\ & &
   \qquad \mbox{} \times 
   j_l(ka_{j}) \frac{h^{(1)}_{l_1}(kR_{jj'})}{h^{(1)}_{l'}(ka_{j'})}
        Y_{l_1,m-\tilde{m}}(\hat{R}^{(j)}_{jj'}) D^{l'}_{m'\tilde{m}}(j,j')
    \; .  \nn
\eee
Details are given in App.\,A.  Here, $j_l(z)$ and $h^{(1)}_l(z)$ are
spherical Bessel functions and Hankel functions of first kind,
respectively~\cite{Abramovitz}.  The vectors $\vec s_j$ are measured
in the global coordinate system attached to the large 
sphere of integration and point from its origin to the 
center of the $j^{\,\rm th}$
sphere (which is of radius $a_j$), $s_j$ is its modulus, and $\hat
s_j$ the corresponding unit vector.  $\vec R_{j j'}$ is the vector
from the center of the sphere $j$ to the center of $j'$ as measured in
the local coordinate system attached to the $j^{\,\rm th}$ sphere,
$R_{j j'}$ is its modulus and $\hat R_{j j'}$ the corresponding unit
vector.  $D^{l}_{m,m'}(gl,j')$ and $D^{l}_{m,m'}(j,j')$ are the
rotational matrices which transform the local coordinate system of
sphere $j'$ to the global coordinate system and to the local coordinate
system of sphere $j$, respectively. Finally, we use the definition of
Ref.\,\cite{Landolt} for the $3j$-symbols.

{\bf Second case:} $\vec{r}{\,}' \in V$, $r'$ large 
\\ \nopagebreak
Because of Green's theorem~(\ref{Green}) we have an explicit expression
for $\psi^k_{lm}(\vec{r})$ 
which yields inserted into the asymptotic expansion~(\ref{Def_S})
an explicit formula for the scattering matrix
\be   \label{S_2}
     \mbf{S} = \mbf{1} - i \mbf{A}^{j} \cdot \mbf{D}^{j} \; ,
\ee
where the elements of the matrix $\mbf{D}^{j}$ are found to be 
\bee    \label{D_new}
   \mbf{D}^j_{lm,l'm'} & = & \frac{ka_j^2}{\sqrt{\pi}} \sum_{l_1=0}^\infty
    \sum_{\tilde{m}=-l}^l (-1)^{\tilde{m}+l_1} i^{l_1-l} \\ & &
   \quad \mbox{} \times\sqrt{(2l+1)(2l'+1)(2l_1+1)}\dreij{l'}{l_1}{l}{0}{0}{0}
    \dreij{l'}{l_1}{l}{m'}{\tilde{m}-m'}{-\tilde{m}}   \nn  \\ & &
   \quad \mbox{} \times j_l(ka_j) j_{l_1}(ks_j) D^{l}_{\tilde{m}m}(j,gl)
             Y_{l_1,\tilde{m}-m'}(\hat{s}_j) \; .   \nn
\eee
Again, details are given in App.\,A.
Here, the rotational matrix $D^{l}_{m m'}(j,gl)$ transforms the global 
coordinate system into the local coordinate system of the $j^{\,\rm th}$ 
sphere. The gradient
matrix $\mbf{A}^j$ appears again because of the boundary integrals on 
the scatterer surfaces. 

Hence, combining the results of both cases, we can use Eq.~(\ref{CAM}) 
to eliminate the matrix
$\mbf A^j$ from Eq.\,(\ref{S_2})~\cite{GaRiqm}. We finally arrive at
\be \label{SMCD}
\mbox{\bf S}  =  \mbox{\bf 1} - i\mbf{C}^j \cdot
                              \left(\mbf {M}^{-1}\right)^{j j'}
                                \cdot  \mbf{D}^{j'}  
\ee
which is expressed in the global coordinate system.
In order to shorten the notation we will suppress the labels $j$ and $j'$,
unless otherwise specified.

As mentioned above the effects of the scattering on the $N$ spheres
are described by the product $-i \mbf{C} \cdot \mbf{M}^{-1} \cdot
\mbf{D}$ which is of course the on-shell {\bf T}-matrix.  Examining
the obtained formulae we see that {\bf M} depends only on the relative
positions of the scatterers and that it contains all the information
about the geometry of the entire scattering system.  It does not
depend on the choice of the center of the large sphere in the
integration volume and on the orientation of the corresponding global
coordinate system.  For these reasons we call {\bf M} a characteristic
matrix. It is of KKR-type~\cite{Berry_KKR,KKR}.  In contrast, the
matrices {\bf C} and {\bf D} depend on this choice
and furthermore do not contain any information about the relative
positions of the $N$ scatterers.  We may therefore conclude that the
coherent multi-sphere part of the scattering is contained in {\bf M} while the
single sphere aspects are  contained in~{\bf C} and~{\bf D}.

If one would like to construct {\bf S} explicitly from
Eq.\,(\ref{SMCD}), it would be necessary to find the inverse of the
infinite matrix {\bf M}, which is a nontrivial task.  But if one is
only interested in spectral properties like scattering resonances it
suffices to look for the poles of the determinant of the {\bf
S}-matrix~\cite{LlSm,Lloyd,Krein,Friedel}. The latter can be expressed
in such a way that it does not involve $\mbf{M}^{-1}$ any more (see
the following Section). {}From Eq.\,(\ref{SMCD}) we expect that the
resonances of the coherent part of the scattering are given by the
zeros of the determinant of {\bf M}.

Before focussing on the determinant of {\bf S}, we should have a
closer look at Eq.\,(\ref{CAM}) which is only determined up to a
transformation of the form $\mbf{C'} \equiv \mbf{CE} = \mbf {AME}
\equiv \mbf{AM'}$.  To get an unambiguous definition, we have chosen
such a normalization that in the case of the scattering from only one
sphere we have simply $\mbf{C} = \mbf{A}$. This choice implies
$\mbf{M} = \mbf{1} + \mbf{W}$ where the diagonal entries of {\bf W}
vanish. In Sec.\,4 we sketch how to prove that $\Tr \mbf{W}$ converges
absolutely in every basis so that the determinant of {\bf M} is well
defined~\cite{rs4}. This is very important as $\Det \mbf{M}$ plays a
crucial role in calculating scattering resonances, as we will see now.


\section{Determinant of the Scattering Matrix and
         the Link to the Semiclassical Zeta-Function}

Since we are interested in the scattering resonances it is sufficient
to find the poles of the determinant, $\det\, \mbf{S}$, of the
scattering matrix~\cite{LlSm,Lloyd,Krein,Friedel} as a function of
complex wave number $k$.  Obviously {\bf S} is an infinite matrix,
hence it is a nontrivial task to prove the existence of its
determinant. To keep the discussion transparent, we postpone the
formal proof to Sec.\,4 and anticipate the
mathematical soundness of the calculations to be performed. We work
out an explicit expression for $\det\,\mbf{S}$ in Sec.\,3.1.
In Sec.\,3.2, we conjecture a direct link to the semiclassical
zeta-function.

\subsection{Calculation of the Determinant}

A formal definition of the determinant of an
infinite matrix $\mbf{Q} \equiv \mbf{1} + \mbf{P}$ is given by
\bee    \label{Def_det}
   \det( \mbox{\bf 1} + \mbox{\bf P}) & = & \exp \{ \tr 
[ \ln(\mbox{\bf 1}+\mbox
{\bf P})] \}  \; ,   \\
        \label{Def_ln}
   \ln(\mbox{\bf 1}+\mbox{\bf P}) & = & -\sum_{n=1}^\infty 
\frac{(-1)^n}{n}  
{\mbox{\bf P}}^n   \; .
\eee
Equation (\ref{Def_det}), understood as Taylor-expanded expression (i.e., in 
the cumulant expansion, see below),
is well defined if $\tr\, \mbf{P}$ converges absolutely in every
basis~\cite{rs4}. Formally one gets the following expression for the
determinant of the $N$-sphere {\bf S}-matrix, $\mbf{S}^{(N)}$,

\begin{eqnarray}   \label{detSdef}
  \det {\mbf{S}}^{(N)} 
    & = & \exp\{\tr \ln(\mbox{\bf 1} - i\mbox{\bf C} \cdot {\mbox{\bf
                                M}}^{-1}
                             \cdot \mbox{\bf D}) \}    \nn \\
    & = & \exp\left\{ -\sum_{n=1}^\infty \frac{i^n}{n} 
      \tr[(\mbox{\bf C} \cdot
                    {\mbox{\bf M}}^{-1} \cdot \mbox{\bf D})^n] \right\}   
\nn \\
    & = & \exp\left\{ -\sum_n \frac{i^n}{n} \Tr 
  [({\mbox{\bf M}}^{-1} \cdot
             \mbox{\bf D} \cdot \mbox{\bf C})^n] \right\}  \nn \\
    & = & \exp\{\Tr \ln(\mbox{\bf 1} - i {\mbox{\bf M}}^{-1} \cdot
             \mbox{\bf D} \cdot \mbox{\bf C}) \}  \nn \\
    & = & \Det (\mbox{\bf 1} - i {\mbox{\bf M}}^{-1} \cdot
             \mbox{\bf D} \cdot \mbox{\bf C})  \nn \\
    & = & \Det [ {\mbox{\bf M}}^{-1} \cdot (\mbox{\bf M} - i\mbox{\bf D
} \cdot \mbox{\bf
C})]              \nn \\
            \label{det_S_a}
    & = & \frac{\Det 
    (\mbox{\bf M} - i\mbox{\bf D} \cdot \mbox{\bf C})}{
  \Det    (\mbox{\bf M})} \; .
\end{eqnarray}
Here, we introduced capital case traces and determinants $\Tr \cdots$ 
and $\Det \cdots$ in order to indicate that they refer to matrices labeled
by the index triples $j,l,m$, 
whereas the lower case traces and
determinants act on matrices 
which are just labeled by the (angular momentum) index pairs $l,m$.

With the caveat that there might be  also poles and zeros in the 
{\em numerator}, the  
%
%
resonances can be determined by just looking for the zeros of $\Det
\mbf{M}$ in the complex $k$-plane. In order to get a simpler
expression for $\det \mbf{S}^{(N)}$, which also gives more physical
insight, we calculate the determinant of $\mbf{X} \equiv \mbf{M} - i
\mbf{D} \cdot\mbf{C}$. 
With the knowledge of  Eqs.\,(\ref{C_N_ball_new}) and (\ref{D_new}) one
can calculate the product $\mbf{D} \cdot \mbf{C}$. In order to do this, the
properties of consecutive rotations (changes of coordinate systems) and the
re-coupling of angular momenta via 6-$j$-symbols (see Ref.\,\cite{Landolt} for
the necessary formulae) have to be considered. Here only the result will be
given,
\bee   
   \mbf{D}^{j}_{lm, l''m''} \mbf{C}^{j'}_{l'' m'',l'm'} & = 
 & \frac{2}{i} \frac{j_l(ka_j)}{h_{l}^{(1)}(ka_j)}
        \delta^{jj'} \delta_{ll'} \delta_{mm'} \nn \\
  && \mbox{} +
    (1-\delta^{jj'}) \sum_{l_1=0}^\infty \sum_{M=-l'}^{l'}
      \left(\frac{a_j}{a_{j'}}\right)^2 \frac{2}{i} \sqrt{4\pi} i^{l_1+l'-l}
         (-1)^m        \nn  \\
  && \quad \mbox{} \times   
    \sqrt{(2l+1)(2l'+1)(2l_1+1)} \frac{j_l(ka_j)}{h^{(1)}_{l'}(ka_{j'})}
      j_{l_1}(kR_{jj'}) Y_{l_1,m-M}(\hat{R}^{(j)}_{jj'})  \nn \\
  && \quad \mbox{} \times
    \dreij{l'}{l}{l_1}{0}{0}{0} \dreij{l'}{l}{l_1}{M}{-m}{m-M}
          D^{l'}_{m'M}(j,j') \; .   
 \label{Y}
\eee
It is easy to see that the matrix $\mbf{X} \equiv \mbf{M} - i 
\mbf{D}\cdot\mbf{C}$ 
is given by
\be   \label{X}
   \mbf{X}^{jj'}_{lm,l'm'} = (-1)^{m'-m} \mbf{S}^{(1)}_{lm,l'm'}(j')
     \left(\mbf{M}^{jj'}_{l,-m,l',-m'} \right)^\ast  \; ,
\ee
where $\mbf{M}^\ast$ is a shorthand for $\left( \mbf{M} (k^\ast) \right)^\ast$
and $\mbf{S}^{(1)}(j;k)$ denotes the 1-scatterer {\bf S}-matrix for the
scattering from a single sphere with radius $a_j$, described in a coordinate
system whose origin lies in the center of the sphere, 
\be   \label{S_1ball_new}
   \mbf{S}^{(1)}_{lm,l'm'}(j') 
    = -\frac{h^{(2)}_{l'}(ka_{j'})}{h^{(1)}_{l'}(ka_{j'})}
         \delta_{ll'} \delta_{mm'}  \; .
\ee
The last equation is easily obtained by
a comparison of the exact solution of this integrable problem and the
ansatz (\ref{Def_S}). 
The corresponding determinant, $\det {\mbox{\bf S}}^{(1)}(j;k)$, is of course
independent of the coordinate system. 
We now have a (formal expression for the) 
product-structure for
the determinant of {\bf X},
\be   \label{det_X}
  \Det \mbox{\bf X}(k)  =  \Det ({\mbox{\bf
M}}(k^\ast))^\dagger
     \left(  \prod_{j=1}^N \det {\mbox{\bf S}}^{(1)}(j;k) \right) \; .
\ee
Combining (\ref{det_X}) and (\ref{det_S_a}), we finally obtain
\be    \label{det_S}
   \det {\mbox{\bf S}}^{(N)}(k) =
      \frac{\Det (\mbox{\bf M}(k^\ast))^\dagger}{\Det
\mbox{\bf M}(k)} \left(
  \prod_{j=1}^N \det {\mbox{\bf S}}^{(1)}(j;k) \right)
  \; .
\ee
The determinant splits up into an incoherent part, consisting of the
product of the $N$ determinants of the one-scatterer {\bf S}-matrices,
and a coherent part, given by the ratio of the determinants of the
hermitian conjugate of the characteristic matrix and the determinant
of {\bf M} itself. Equation~(\ref{det_S}) obviously respects the unitarity
of the {\bf S}-matrix, as the
one-scatterer {\bf S}-matrices~(\ref{S_1ball_new}) are by themselves 
unitary because
of $(h^{(1)}_m (ka))^\ast = h^{(2)}_m(k^\ast a)$ 
and as the coherent part of $\det
\mbf{S}^{(N)}$ is manifestly unitary. 
As mentioned, the scattering resonances are given by the zeros of  
$\Det \mbf{M}$. However, this
determinant 
does not only possess zeros but also poles. These poles cancel the
resonance poles of the incoherent part of $\det \mbf{S}^{(N)}$, as
$\Det \mbf{M}$ and the product of the $\det {\mbox{\bf S}}^{(1)}$ both
involve the same number and power of Hankel functions  
$h^{(1)}_l (ka_j)$. The same is
true for the poles of $\Det \mbf{M}^\dagger$ and the zeros of the
product of the $\det {\mbox{\bf S}}^{(1)}$: both involve the same number
and power of Hankel functions $h^{(2)}_l (ka_j)$.

It is clear that $\det \mbf{S}^{(N)}$ does not depend on the choice of
the global coordinate system that was used in the
definition of the {\bf S}-matrix (\ref{Def_S}) and to fix the center
of the large spherical integration volume in Green's theorem
(\ref{Green}).  As we are interested in the coherent part of the
scattering we will deal with $\Det \mbf{M}(k)$ from now on.

The symmetry of the scattering configuration leads to a block-diagonal
form of the matrix {\bf M}. If the symmetry group of the system is
finite, we obtain [see App.\,B for the details of symmetry reductions]
\be    \label{det_S_zerl}
  \det {\mbox{\bf S}}^{(N)}(k) = \left(
         \prod_{j=1}^N \det {\mbox{\bf S}}^{(1)}(j;k) \right)
    \frac{\prod_c 
      (\Det (\tilde{\mbox{\bf M}}_{D_c}(k^\ast))^\dagger)^{d_c}}
        {\prod_c
      (\Det \tilde{\mbox{\bf M}}_{D_c}(k))^{d_c}} \; ,
\ee
where the index $c$ runs over all conjugacy classes and $D_c$ denotes
the $c^{\,\rm th}$ irreducible representation of dimension $d_c$ of
the symmetry group.  The last formula represents the final result of
our formal treatment of scattering systems consisting of $N$ spherical
scatterers. The scattering resonances corresponding to the $c$-th
irreducible representation of the symmetry group are given by the
zeros of $\Det \tilde{\mbox{\bf M}}_{D_c}(k)$ and they are $d_c$-fold
degenerate.

In Sec.\,4 it will be shown that all the formal manipulations
that lead to Eqs.\,(\ref{det_S}) and (\ref{det_S_zerl}), and
especially all the determinants appearing in it, are well defined if
the number $N$ of scatterers is finite and the scatterers do not
overlap nor touch.

\subsection{Connection to the Semiclassical
            Gutwiller-Voros Zeta-Function}

The determinant of the characteristic matrix {\bf M} is understood in terms
of the cumulant expansion~\cite{unspaper,AW_testof}, which can be formally
obtained from Eqs.\,(\ref{Def_det}) and (\ref{Def_ln}),
\bee
  \Det \mbf{M} & = & \exp \left\{ \Tr \log \mbf{M}
  \right\} \, \equiv \, \exp \left\{ \Tr \log (\mbf{1} +  \mbf{W})
  \right\}
\nn  \\          \label{Cumulant_entw}
  & = & 1 +  \, \Tr \mbf{W} - \frac{1}{2} \left[ \Tr (\mbf{W})^2
-         \left( \Tr \mbf{W} \right)^2 \right] + \ldots  \; .
\eee
The first line of the upper equation is only of formal character ---
especially it is not defined at the zeros of $\Det
\mbf{M}$. Nevertheless, it offers an easy way to remember the second
line, which defines the determinant of {\bf M}, if {\bf W} is trace
class (see Ref.\,\cite{unspaper} and references therein).

In a semiclassical description the scattering resonances can be extracted
from the zeros of the Gutzwiller-Voros zeta function, which formally can
be written as~\cite{gutzwiller,voros88}
\bee   \label{zeta}
  Z_{GV} (k;z) & = & \exp \left\{ - \sum_p \sum_{r=1}^\infty \,\frac{1}{r}
     \frac{(z^{n_p} t_p)^r}{ |\det(\mbf{J}_p{}^r - \mbf{1})|^{\frac{1}{2}}}
      \right\}  \\
  t_p & = & \exp \left\{ i \left( \frac{S_p(k)}{\hbar} - \nu_p \frac{\pi}{2}
\right) 
    \right\}  \; . \nn
\eee
Here $\mbf{J}_p$ denotes the monodromy matrix of the $p$-th primitive periodic
orbit of topological length $n_p$, $S_p(k)$ and $\nu_p$ are the corresponding 
classical action and Maslov index, respectively. The sum over $r$ takes into
account the
repeated traversals of the primitive periodic orbits. 
In the above expression, $z$ is a book-keeping variable and the
Gutzwiller-Voros zeta function has to be evaluated at $z=1$. This
expression is only of formal character, it has to be regulated. Expanding 
(\ref{zeta}) in powers of $z$ the  curvature expansion is obtained,
\bee  \label{curv}
 Z_{GV} (k;z) & = & 1 - z \sum_p \,\delta_{n_p,1}\, 
  \frac{t_p}{\Gamma_{p,r=1}} - 
  \frac{z^2}{2} c_2 + \ldots   \\
 c_2 & = & 2 \sum_p \sum_r \,\delta_{n_pr,2} \,\frac{1}{r}\,
\frac{(t_p)^r}{\Gamma_{p,r}} - \sum_{p,p'}\, \delta_{n_p,1}\, 
  \delta_{n_{p'},1}\,
   \frac{t_p t_{p'}}{\Gamma_{p,r=1}
 \Gamma_{p',r=1}}   \label{c_2} \\
 \Gamma_{p,r} & \equiv & |\det(\mbf{J}_p{}^r - \mbf{1})|^{\frac{1}{2}} \, = \,
|\Lambda_{1,p} \Lambda_{2,p}|^{\frac{1}{2}} (1-\Lambda_{1,p}{}^{-r})
(1-\Lambda_{2,p}{}^{-r})  \; . \nn
\eee

The $n$-th curvature $c_n$ contains all periodic orbits up to the topological
length $n$. As can be seen from Eq.\,(\ref{c_2}), the contribution of orbits of
topological length $n$~($>1$) gets reduced by pseudo-orbits, composed of shorter
periodic orbits,  of the same total
length.  
Equations (\ref{zeta}) and (\ref{curv}) (except for the last equality in the
last line) are valid for two-dimensional $N$-disk systems as well as
for three-dimensional $N$-sphere systems. In contrast to the well known
two-dimensional case, in three dimensions the monodromy matrix, as being a $4
\times 4$-matrix, has two leading eigenvalues
$|\Lambda_1|,|\Lambda_2|>1$~\cite{gutzwiller}.  For
a generic periodic orbit in an $N$-sphere system the two leading eigenvalues are
different. The most important exception is the only periodic orbit of the
two-sphere-system, here $\Lambda_1$ and $\Lambda_2$ coincide.
In $N$-sphere systems which have a two-dimensional analogue -- this means that 
the centers of the  $N$
spheres are located all in one plane -- one of the leading eigenvalues 
is given by the
leading eigenvalue in the corresponding $N$-disk system while the other takes
into account the instability of the periodic orbit against perturbations
perpendicular to the plane. The lengths and Maslov indices of the periodic
orbits in these $N$-sphere systems 
coincide with the corresponding quantities in
the analogous $N$-disk system. 
As in the two-dimensional $N$-disk systems, orbits
which are situated on the boundary of the fundamental domain require a special
treatment~\cite{lauritzen,Cv_Eck}.

It is easy to see that the series of Eq.\,(\ref{Cumulant_entw})  has the same
structure as the curvature expansion of the
semiclassical Gutz\-willer-Voros zeta function~(\ref{curv}).
Furthermore, the quantum-mechanical description of $N$-sphere 
scattering systems is analogous
to the quantum-mechanical description of two-dimensional $N$-disk systems. In
the treatment of $N$-disk systems a direct link between the quantum-mechanical
and semiclassical descriptions has been established~\cite{unspaper}. On the
quantum-mechanical side this link is based on~(\ref{Cumulant_entw}). Therefore
we conjecture a similar direct link between quantum mechanics and
semiclassics in $N$-sphere systems:
The semiclassical limit of the cumulant expansion in the Plemej-Smithies form
(see Ref.\cite{rs4})
of $\Det \mbf{M}$ is given by the curvature regulated Gutzwiller-Voros zeta 
function plus diffractional
corrections, 
\bee   \label{link}
    \Det {\bf M}(k) & \semiclass &  
 {\widetilde Z}_{GV} (k)|_{\rm curv.\,reg.} \\ 
  \Tr(\mbf{W}^n(k)) & \semiclass & (-1)^n\sum_p\, \delta_{n_pr,n}\, n_p 
    \frac{t_p(k){}^r}{\Gamma_{p,r}} + \mbox{diffractional corrections.} \nn
\eee 
Hence, the
semiclassical limit of the $n$-th cumulant is given by the $n$-th
curvature order in the Gutzwiller-Voros zeta function plus diffractional
corrections. 
This direct link, which connects the quantum-mechanical and semiclassical
descriptions, is valid for all complex $k$ and not only for the
isolated scattering resonance poles.
We call it a direct link because it is not based on concepts characteristic
to bound-state problems. So it is not based on any asymptotic limit 
of spectral densities (see Refs.\,\cite{Pinball} and \cite{BaBl}):
\be
 \lim_{b\to \infty} \left (
 N^{(N)}(k;b)-N^{(0)}(k;b) \right )
 =\frac{1}{2\pi} {\rm Im} \Tr \ln {\bf S}(k)
 \label{friedel_sum}
 \; ,
\ee
where $N^{(N)}(k;b)$ and
$N^{(0)}(k;b)$
are the
integrated spectral densities belonging to two
spherical bound systems, both of the same radius $b$, where one
encircles the scattering region whereas the other doesn't 
(see Ref.\,\cite{unspaper}
for a detailed discussion in the analogous two-dimensional $N$-disk systems).
In fact, this formula is only correct in the double limit $\lim_{\epsilon \to
0} \lim_{b\to \infty}$ where a small positive imaginary part $i\epsilon$ 
has to be added
to the wave number, before the limit $b\to \infty$ can be taken.

In Ref.\,\cite{GePr-FiGePr} a structure similar to 
that of Eq.\,(\ref{Cumulant_entw}) of 
the characteristic determinant
for generic two-dimensional systems was derived in a semiclassical
description under the Fredholm theory.

\section{Justification of the Previous
         Calculations}

The derivation of the expression for the {\bf S}-matrix (\ref{SMCD})
and the derivation of its determinant in Sec.\,3 are of a purely
formal character since all the matrices involved are of infinite size. 
In this Section, we show that the calculations of the previous
Section are mathematically sound. This discussion is of considerable
conceptual importance. However, those readers who are mainly interested 
in the numerical results of our study are adviced to skip the
present Section. 

In the proofs that all the performed operations are well defined the so-called
trace-class and Hilbert-Schmidt operators~\cite{rs4} play a central
role. Trace class operators are those, in general, non-hermitian operators
of a separable Hilbert-Space which have an absolutely
convergent trace in every orthonormal basis. An operator {\bf B}
belongs to the Hilbert-Schmidt class if $\mbf{B}^\dagger \mbf{B}$ is 
trace-class. Here we will not present all the proofs in detail. 
We refer to~\cite{unspaper} where
the corresponding problem in the two-dimensional $N$-disk scattering systems is
treated. In that reference the most important properties of 
trace-class and Hilbert-Schmidt operators are listed:
(i) any trace-class operator can be represented as the product of two
Hilbert-Schmidt operators and any such product is trace-class; 
(ii) an operator $\mbf{B}$ is already 
Hilbert-Schmidt,
if the trace of $\mbf{B}^\dagger \mbf{B}$ is absolutely convergent in just 
one orthonormal basis; (iii)
the 
linear combination of a finite number of trace-class operators is again
trace-class; (iv) the adjoint of a trace-class operator is again trace-class;
(v) the product of two Hilbert-Schmidt
operators or of a trace-class and a bounded operator is trace-class and
commutes under the trace; (vi) if $\mbf{B}$ is trace-class, the determinant
$\det(\mbf{1}+z \mbf{B})$ exists and is an entire function of $z$; (vii) the
determinant is invariant under unitary transformations.

In close analogy to the results of Ref.\,\cite{unspaper} 
the following steps can be proven -- 
provided that $N$ is finite and that the spheres do
not touch nor overlap: 
\begin{description}
\item[(a)\ ]
$\mbf{D}^j$ is a trace-class matrix for all complex $k$.  Also
$\mbf{C}^j$ is a trace-class matrix except at the isolated zeros of
$h_l^{(1)}(ka_j)$, where $l$ is a nonnegative integer and $j = 1,
\ldots ,N$. (One very simple 
way to prove this is to transform $\mbf{D}^j$ and
$\mbf{C}^j$ into the eigenbasis of the $j^{\,\rm th}$ 
sphere, see Eq.(\ref{S_1ball_new}). In that
eigenbasis both matrices become diagonal and the trace-class property can
easily be checked by summing up the moduli of their eigenvalues. In
the same way it can be checked that the one-sphere {\bf T}-matrix  
$\mbf{S}^{(1)}(j;k)- {\bf 1}$ is trace-class, too.)
\item[(b)\ ]
Therefore the product $\mbf {D}^j \mbf{C}^{j'}$ 
is of trace-class as long as $N$ is finite and $\mbf{C}^{j'}$ is trace-class.
\item[(c)\ ]
$\mbf{M}^{j j'} - \delta^{jj'}\mbf{1} = \mbf{W}^{jj'}$ 
is trace-class, except at the same 
$k$-values mentioned in~{\bf (a)}. This can be proved by rewriting 
$\mbf{W}^{jj'}$ as the product of two matrices, $ \mbf{G}^{jj'}$ and a
diagonal matrix
$\mbf{H}^{jj'}$, which both can be shown --- as in
Ref.\cite{unspaper} --- to be Hilbert-Schmidt matrices:
\bee       \label{E_ball}
   \mbf{G}^{jj'}_{lm,l'm'} & = & (1-\delta^{jj'}) 
          \left( \frac{a_j}{a_{j'}} \right)^2
               \sqrt{4\pi} \frac{(-1)^{m}}{(2l'+1)^{\threehalf}} 
          \,\frac{j_l (k a_j)} 
          {\sqrt{h^{(1)}_{2l'}(k\alpha a_{j'})}} \nn \\
  && \quad \mbox{}\times
      \sum_{l_1=0}^\infty \sum_{\tilde{m}=-l'}^{l'} i^{l_1+l'-l} 
      \sqrt{(2l+1)(2l_1+1)(2l'+1)}\nn \\
  && \quad \mbox{} \times
     \dreij{l_1}{l'}{l}{0}{0}{0}
        \dreij{l_1}{l'}{l}{m-\tilde{m}}{\tilde{m}}{-m}    \nn \\ 
  && \quad \mbox{} \times 
     h^{(1)}_{l_1}(kR_{jj'})
        Y_{l_1,m-\tilde{m}}(\hat{R}^{(j)}_{jj'}) D^{l'}_{m'\tilde{m}}(j,j')
\eee
and 
\be
   \mbf{H}^{j'j''}_{l'm',l''m''}  =  \delta^{j'j''}\delta_{l' l''} 
  \delta_{m' m''} 
            \, (2 l'+1)^\threehalf \frac{\sqrt{h^{(1)}_{2l'}(k \alpha a_{j'})}}
       {h^{(1)}_{l'}(k a_{j'})} \; ,
\ee
where $\alpha > 2$. This inequality is the reason that our proofs
exclude the case of touching spheres. In fact, the geometry of the
$N$ scatters must be such that the inequality  
$ \frac{\alpha}{2}a_j + a_{j'} < R_{j j'}$ is valid for all pairs 
$j$ and $j'$. 
\item[(d)\ ]
Therefore $\mbf{M}$ is bounded.
\item[(e)\ ]
$\mbf{M}$ is invertible everywhere 
where $\Det \mbf{M}$ is defined 
and nonzero (which excludes a countable number
of isolated points in the lower
$k$-plane). Especially, $\mbf{M}$ is invertible on the real $k$ axis.
Therefore,  the matrix
${\bf M}^{-1}$ is bounded on the real $k$ axis as well.
\item[(f)\ ]
${\bf C}^j ({\bf M}^{-1})^{jj'} {\bf D}^{j'}$, 
${\bf M}^{-1} {\bf D C}$,
are all of trace-class (except at the isolated points 
mentioned in {\bf (a)} and
{\bf (e)} as they are the product of a (finite number of) 
bounded and
trace-class matrices, and 
$\tr[({\bf C}^j ({\bf M}^{-1})^{j j'} {\bf D}^{j'})^n]=
\Tr[( {\bf M}^{-1} {\bf D C})^n ]$ exists (note
the matrices on the l.h.s.\ are
labeled by the index pairs $l,m$, whereas the ones on the r.h.s.\ are labeled
by the index triples $j,l,m$).  
\item[(g)\ ]
 ${\bf M}-i{\bf DC}-{\bf 1}$ is of  trace-class because of {\bf (b)}
and {\bf (c)} and the rule that the sum of two trace-class matrices is
again trace-class. 
\end{description}

The properties {\bf (a)} -- {\bf (g)} ensure that the derivation of
Eq.\,(\ref{det_S_a}) starting from~(\ref{SMCD}) is correct.  Because of
{\bf (e)} formula~(\ref{SMCD}) makes sense (also on the real
$k$-axis).  Under the assumption that the matrix {\bf A}, which
characterizes the gradient of the wave function on the surfaces of the
scatterers (see Eq.\,(\ref{RB})), is bounded,
the 
determinant of {\bf S} on the basis of (\ref{S_2}) is defined.
Now it is easy to see (using property {\bf (a)} and {\bf (c)}) that 
all determinants appearing in the formula
(\ref{det_S}) exists and that the unitary transformations leading to 
(\ref{det_S_zerl}) are justified.


\section{Numerical Results}

We have calculated the scattering resonances of the three simplest and
most symmetrical $N$-sphere systems.  These systems consist of two,
three and four hard spheres, respectively, which all have the {\em
same} radius $a_j\equiv a$ and the {\em same} center-to-center
separation $R_{ij}\equiv R$: two non-touching spheres (two spheres),
three spheres at the corners of an equilateral triangle (three
spheres) and four spheres at the corners of a regular tetrahedron
(four spheres). In all cases we have chosen the ratio $R/a$ to have
the fixed value 6 in order to be able to compare with older two and
three-disk calculations.

The quantum-mechanical resonances have been calculated as the zeros of
the determinant of the characteristic matrix {\bf M} in a finite basis
corresponding to angular momenta from $l=0$ up to a certain maximum
value which depends on the wave number $k$.  We checked the accuracy
of the results against a further enhancement of the basis.  This
method is applicable, as the trace class property of $\mbf{M} -
\mbf{1}$ guarantees the existence of the limit.  The zeros in the
complex $k$-plane were determined with the help of a Newton-Raphson
routine.

The semiclassical resonances have been calculated as the zeros of the
curvature regulated Gutzwiller-Voros zeta function without
diffractional corrections. In the two-sphere and three-sphere systems
the zeta function can be easily constructed under the techniques
described in Sec.\,3, with the input from the corresponding two-disk
and three-disk systems, respectively. The four-sphere-system is the
simplest $N$-sphere scattering system which does not have a
two-dimensional analogue. In this case, only the three fundamental
periodic orbits of topological length 1 have been determined so far.

In Sec.\,5.1, we discuss the quantum mechanically calculated resonances
of the $N$-sphere systems. They are compared to those of the two-dimensional
$N$-disk systems in Sec.\,5.2. In Sec.\,5.3, we compare the quantum
mechanical and the semiclassical results for the $N$-sphere systems.

\subsection{Quantum Mechanically Calculated Resonances}

In Secs.\,5.1.1, 5.1.2 and 5.1.3 we discuss the two-, three and
four-sphere systems, respectively.

\subsubsection{Two Spheres}

The scatterer configuration is shown in Fig.\,1. This system has 
a continuous
symmetry,  the rotational symmetry about the axis which joins the
centers of the spheres. Hence, the corresponding 
symmetry group, $D_{\infty h}$, is infinite and
has 4 one-dimensional irreducible representations and an
infinite number of two-dimensional ones~\cite{Hammermesh}. {}From the 4
one-dimensional representations only two are present due to the lack of any
inner structure of the two spherical scatterers. 

%
%
%
%
%
%
  
The blocks of the symmetry reduced {\bf M}-matrix are given by: 

\begin{samepage}
one-dimensional representations (with $m=0$):
\begin{eqnarray}
           \label{M_2ball_eindim}
   \tilde{\mbf{M}}_{ll'}(D) & = & \delta_{ll'}   
+ (-1)^c (-1)^{l'} i^{l'-l} \frac{j_l(ka)}{h^{(1)}_{l'}(ka)}
        \sqrt{(2l+1)(2l'+1)}     \nn \\  & &
   \qquad\ \ \mbox{} \times \sum_{l_1=0}^\infty i^{l_1} (2l_1+1) \left(
     \dreij{l_1}{l'}{l}{0}{0}{0} \right)^2 h_{l_1}^{(1)}(kR)  \nn \\
   c & = & \cases{ 0 & for $D_c = A_{1g}$    \cr
                 1 & for $D_c = A_{1u}$      }  \nn
\end{eqnarray}
\end{samepage}

\begin{samepage}
two-dimensional representations:
\begin{eqnarray}
     \label{M_2ball_zweidim}
   \tilde{\mbf{M}}_{ll'}(D) & = & \delta_{ll'}  
  +  (-1)^c\, (-1)^{l'}\, i^{l'-l} \frac{j_l(ka)}{h^{(1)}_{l'}(ka)}
        \sqrt{(2l+1)(2l'+1)}  \nn \\  & &
   \qquad\ \ \mbox{} \times 
 \sum_{l_1=0}^\infty i^{l_1} \,(2l_1+1) \dreij{l_1}{l'}{l}{0}{0}{0}
      \dreij{l_1}{l'}{l}{0}{m}{-m} h_{l_1}^{(1)}(kR)  \nn  \\
    c & = & \cases{ 0 & for $D_c = E_{mg}$    \cr
                    1 & for $D_c = E_{mu}$      }     \nn \\
    l,l' & \ge & m \; > \; 0  \; , \; \mbox{$m$ fixed}. \nn
\end{eqnarray}
\end{samepage}
 
The degeneracy-degree  of the resonances coincides with the
dimensionality of the representation. As a consequence of the continuous
rotational symmetry, $|m|$ is a good quantum number and there are two
irreducible representations corresponding to each value of $|m|$.

The numerically calculated resonances of the two-sphere-system are shown in
Figs.\,4-6. The leading resonances of Fig.\,4
form a regular structure which we call a Gutzwiller-band. The suppressed
resonances do not build up such a regular structure; we say they lie in a
diffraction-band. This terminology is chosen in analogy to the similar results
found in two-dimensional problems (scattering from $N$ hard
disks)~\cite{GaRiqm,unspaper,AW_testof,vwr_prl,vwr_jap}. 
The resonances corresponding to
increasing $|m|$ 
become increasingly suppressed.
This is due to the centrifugal barrier, which in  cylindrical coordinates
reads $m^2/{r^2}$.

In Fig.\,5 we see that the leading resonances of the $|m|=1$
representations for small real parts of $k$ lie in the diffraction band.
For real parts of $k$ bigger than approximately $10/a$ two Gutzwiller 
bands build up, which are shifted by
half a spacing in Re~$k$ but coincide in their values of Im~$k$.

\subsubsection{Three Spheres}

The scatterer configuration of the three-sphere-system is shown in
fig~2. The symmetry group $D_{3h}$ consists of 12 elements. It has 
four one-dimensional and two two-dimensional representations~\cite{Hammermesh}.

%
%
%
%
%
%
%
%
%
%
%

The symmetry reduced expressions for the distinct blocks of the {\bf M}-matrix
are given as follows.

\begin{samepage}
There are four one-dimensional representations ($0 \le m, \, m'$):
\bee   \label{M_A_1p}
  \tilde{\mbf{M}}_{lm,l'm'}({A_1}') & = & \cases{
        0 & ,\ \ $(l+m)$ or $(l'+m')$ odd  \cr
        \mbf{E}_{lm,l'm'}(\alpha = 0) & ,\ \ otherwise }   \\
       \label{M_A_1pp}
  \tilde{\mbf{M}}_{lm,l'm'}({A_1}'') & = & \cases{
        0 & ,\ \ $(l+m)$ or $(l'+m')$ even  \cr
        \mbf{E}_{lm,l'm'}(\alpha = 1) & ,\ \ otherwise }   \\
       \label{M_A_2p}
  \tilde{\mbf{M}}_{lm,l'm'}({A_2}') & = & \cases{
        0 & ,\ \ $(l+m)$ or $(l'+m')$ odd  \cr
        \mbf{E}_{lm,l'm'}(\alpha = 1) & ,\ \ otherwise }   \\
       \label{M_A_2pp}
  \tilde{\mbf{M}}_{lm,l'm'}({A_2}'') & = & \cases{
        0 & ,\ \ $(l+m)$ or $(l'+m')$ even  \cr
        \mbf{E}_{lm,l'm'}(\alpha = 0) & ,\ \ otherwise }   \\ \nn 
\eee
\end{samepage}

\begin{samepage}
\bee
  \mbf{E}_{lm,l'm'}(\alpha) & = & \delta_{ll'} \delta_{mm'}  \nn \\ & &
  \quad \mbox{}+ 
 2 g_m g_{m'} \sqrt{4\pi} (-1)^m i^{l'-l} \frac{j_l(ka)}{h^{(1)}_{l'}(ka)}
      \nn \\ & &
   \quad\ \mbox{}\times 
    \sum_{l_1=0}^\infty i^{l_1} h^{(1)}_{l_1}(kR)
     \sqrt{(2l+1)(2l'+1)(2l_1+1)} \dreij{l_1}{l'}{l}{0}{0}{0} \; 
             \nn \\ & &
   \quad\ \mbox{}\times
   \Bigg\{ \dreij{l_1}{l'}{l}{m-m'}{m'}{-m}
      \cos\left( \frac{\pi}{6}(5m-m') \right) Y_{l,m-m'}(\frac{\pi}{2},0) 
             \nn \\  & &
     \quad\ \ \mbox{} + (-1)^{\alpha+m'} \dreij{l_1}{l'}{l}{m+m'}{-m'}{-m}
      \cos\left( \frac{\pi}{6}(5m+m') \right) Y_{l,m+m'}(\frac{\pi}{2},0)
         \Bigg\}     \nn \\
   g_m & = & \cases{
             1 & ,\ \ $m>0$ \cr
             \frac{1}{\sqrt{2}} & ,\ \ $m=0$. }     \nn
\eee
\end{samepage}

\begin{samepage}
In addition, there are two two-dimensional representations ($m, \, m'$ integer):
\bee    \label{M_E_p}
  \tilde{\mbf{M}}_{lm,l'm'}(E') & = & \cases{
          0 & ,\ \ $(l+m)$ or $(l'+m')$ odd  \cr
        \mbf{F}_{lm,l'm'} & ,\ \ otherwise }   \\
        \label{M_E_pp}
  \tilde{\mbf{M}}_{lm,l'm'}(E'') & = & \cases{
          0 & ,\ \ $(l+m)$ or $(l'+m')$ even  \cr
        \mbf{F}_{lm,l'm'} & ,\ \ otherwise }   
\eee
\end{samepage}

\begin{samepage}
with
\bee
  \mbf{F}_{lm,l'm'} & = & \delta_{ll'} \delta_{mm'}  \nn \\ & &
    \quad \mbox{}+
     2 \sqrt{4\pi} i^{l'-l} (-1)^m \frac{j_l(ka)}{h^{(1)}_{l'}(ka)}
             \nn \\ & &
    \quad\ \mbox{} \times
     \sum_{l_1=0}^\infty i^{l_1} \sqrt{(2l+1)(2l'+1)(2l_1+1)}
      \dreij{l_1}{l'}{l}{0}{0}{0} \dreij{l_1}{l'}{l}{m-m'}{m'}{-m}  \nn \\ & &
    \quad\ \mbox{} \times
     h^{(1)}_{l_1}(kR) Y_{l_1,m-m'}(\frac{\pi}{2},0)
       \cos\left( \frac{\pi}{6} (5m-m'-4) \right)   \; .  \nn
\eee
\end{samepage}

The numerically calculated resonances of the $A_1{}'$ and $A_1{}''$
symmetry classes are shown in
Fig.\,7. The $A_1{}''$ resonances
are suppressed compared with the $A_1{}'$ resonances, because
the wave function has to vanish in the plane of the triangle due to the
symmetry (see character table of $D_{3h}$ in Ref.\,\cite{Hammermesh}). This is
the plane which contains all geometric periodic orbits of the 
three-sphere-system,
which play the dominant role in a semiclassical treatment in the considered
wave number 
regime. The resonances of the $A_2{}'$, $A_2{}''$ and $E'$, $E''$ symmetry
classes show a similar behaviour.

\subsubsection{Four Spheres}

The scatterer configuration of the four-sphere-system is shown in
fig~3. Compared with the two- and three-sphere-systems 
this is the first
genuine three-dimensional scattering system since the scatterers (and therefore
the set of all geometric periodic orbits) do not lie in a plane.

%
%
%
%
%
%
%
%
%
%
%
%
%
%
%
%
%
%
%
%
%
%
%
%
%
%

The symmetry group $T_d$ has 24 elements. It has 2 one-dimensional, 1
two-dimensional and 2 three-dimensional irreducible
representations~\cite{Hammermesh}. Here only the one-dimensional
representations are considered.  The corresponding blocks of the
symmetry reduced {\bf M}-matrix are given by:
  
\bee    \label{M_A_4ball}
  \tilde{\mbf{M}}_{lm,l'm'}(D) & = & \delta_{ll'} \delta_{mm'}  \\ & &
     \quad \mbox{} +
 \frac{3}{2} \sqrt{4\pi} i^{l'-l} \frac{j_l(ka)}{h^{(1)}_{l'}(ka)}
           g_m g_{m'} \nn \\ & &
   \qquad \mbox{} \times
     \sum_{\tilde{l} =0}^\infty \sum_{M=-l'}^{l'} i^{\tilde{l}}
      h^{(1)}_{\tilde{l}}(kR) \sqrt{(2l+1)(2l'+1)(2\tilde{l}+1)}
      \dreij{\tilde{l}}{l'}{l}{0}{0}{0}  \;  \nn \\ & &
   \qquad \mbox{} \times
     (-1)^M \left( d^{l'}_{m'M}(\beta_0) +
          (-1)^{Q + m'} d^{l'}_{-m',M}(\beta_0)
            \right) \;  \nn \\ & &
   \qquad \mbox{} \times
     \Bigg[ (-1)^m Y_{\tilde{l},m-M}(\theta_0,0)
       \dreij{\tilde{l}}{l'}{l}{m-M}{M}{-m}  \nn \\ & &
      \qquad\quad \mbox{} +
    (-1)^Q Y_{\tilde{l},-m-M}(\theta_0,0)
       \dreij{\tilde{l}}{l'}{l}{-m-M}{M}{m} \Bigg]  \nn  \\
%
  Q & = & \cases{  0 & for $D = A_1$  \cr
                   1 & for $D = A_2$  }     \nn   \\
  g_m & = & \cases{ \sqrt{2} & for $m=0$  \cr
                           1 & for $m=3,6,9,\ldots,l$ \cr
                           0 & otherwise  }  \nn \\
  d^j_{mm'}(\beta) & \equiv & \bra{jm} e^{-i \beta J_y} \ket{jm'} \nn \\
  \cos(\theta_0) & = & - \frac{2}{\sqrt{6}} \qquad , \qquad
  \sin(\theta_0) \; = \; \frac{1}{\sqrt{3}}  \nn  \\
  \cos(\beta_0) & = & - \frac{1}{3}  \qquad , \qquad
  \sin(\beta_0) \; = \; \frac{2}{3} \sqrt{2}  \nn
%
\eee
The numerically calculated $A_1$-resonances are shown in Fig.\,8.

A comparison of the first leading Gutzwiller resonances of the two-,
three- and four-sphere-systems shows that the spacing in the real part
of $k$ in the resonance band slightly decreases if one looks first at
the two-sphere, then at the three-sphere and finally at the
four-sphere system. This spacing is governed by the inverse length of
the (averaged) periodic orbits of topological length one in the
fundamental domain.  In the two-sphere case there exists only one
periodic orbit.  The three-sphere system has in addition to this orbit
one further fundamental periodic orbit, which in the global domain
corresponds to an equilateral triangle spanned inbetween the three
scatterers.  The length in the fundamental domain of this second
periodic orbit is only slightly bigger than that of the two-sphere
orbit.  In the four-sphere system there are three fundamental periodic
orbits: The two orbits of the three-sphere system and an additional
orbit which touches all four spheres in the global domain and which
again it slightly bigger than the other two when measured in the
fundamental domain. Thus the average length of the fundamental orbits
increases with the increasing number of spheres.

As far as the imaginary parts of the first leading Gutzwiller
resonances of the two-, three- and four-sphere-systems are concerned,
we see that the two-sphere resonances are slightly more suppressed
than the three-sphere resonances, which are in turn slightly
suppressed as compared to the four-sphere-resonances. This is due to
the fact that the addition of one further sphere increases the
probability that the particle is rescattered to the other spheres and
therefore trapped for a longer time in the scattering region.

\subsection{Comparison of the Quantum Mechanically Calculated 
            Resonances of $N$-Sphere and $N$-Disk Systems}

If the centers of the spheres of a given $N$-sphere configuration lie
all in one plane, there exists an analogous two-dimensional $N$-disk
scattering system. In this case it makes sense to compare directly
these analogue systems, although they differ in their dimensionality:
they have in common the entire set of classically allowed periodic
orbits which play the dominant role in a semiclassical description in
the wave number regime considered
here~\cite{GaRiskl,AW_testof,RoWhWi}.

In Fig.\,9 the resonances of the totally symmetric
representations of the two-sphere and two-disk systems are shown. In
both cases the leading resonances lie in Gutzwiller bands. Suppressed
resonances form diffraction bands. However, in the two-disk system a
sub-leading Gutzwiller band seems to appear for larger wave numbers
(${\rm Re}\, k > 15/a$).  The leading resonances in the two-sphere and
two-disk systems have the same real part, but the whole two-sphere
band is shifted down into the negative complex $k$ plane, because the
only existing geometrical periodic orbit is more instable in the
three-dimensional case than in the two-dimensional one.  In the
semiclassical formulae the higher instability is easily explained by
the existence of a second leading eigenvalue of the monodromy matrix
(see Sec.\,3.2). There is no such coincidence in the real parts of the
resonances of the diffraction bands. This behaviour is also expected
from a semiclassical point of view as the diffractional orbits in both
cases act on different manifolds.

In Fig.\,10 the resonances of the totally symmetric
representations of the three-sphere and three-disk systems are
shown. In both cases the leading resonances lie in Gutzwiller
bands. Suppressed resonances form diffraction bands.  As in the
two-scatterer systems, we find a good agreement between the real parts
of the sphere and disk resonances (which is getting better for larger
wave numbers) while the whole three-sphere Gutzwiller band is shifted
to smaller imaginary parts by the same amount as in the two-scatterer
case. The same behaviour is observed for the two-dimensional
representations $E'$ (three-sphere) and $E$ (three-disk),
respectively.

In summary we see, without using the results of any semiclassical
calculation, that the leading resonances in the analogous two- and
three-dimensional $N$-scatterer systems are dominated by the
contribution of the geometric periodic orbits. In contrast, the
suppressed resonances show no one-to-one correspondence between the
disk and sphere resonances. So the conclusion, which so far is only
based on the presented quantum-mechanical data, is that the suppressed
resonances are due to diffractional effects, which should be different
in two and three dimensions.

\subsection{Comparison of Quantum Mechanically and Semiclassically 
            Calculated Resonances in $N$-Sphere Systems}

As described in Sec.\,3.2, it is possible to transfer the well
developed techniques of calculating scattering resonances under the
periodic orbit approximation with only classically allowed
orbits~\cite{GaRiskl,Pinball,AW_testof} from $N$-disk systems to
scattering problems with $N$ hard spheres in three dimensions.

In particular, the semiclassical resonances of the two-sphere-system 
are given by
the zeros of the following spectral zeta function,
\be   \label{Z_2-ball}
  Z(k) = \prod_{c=0}^\infty Z_{c=|m|}^{g/u} (k)
    = \prod_{m=-\infty}^{\infty} \prod_{n=0}^\infty \left(
    1 \pm \frac{e^{ik(R-2a)}}{|\Lambda_0| \Lambda_0^{|m|+2n}} \right) \; ,
\ee
where the $\pm$ sign refers to the $g/u$ representations and $Z_{|m|}$
contains all terms of a given $|m|$ in the last expression.
$\Lambda_0$ is the leading eigenvalue of the monodromy matrices of the 2-disk
and the two-sphere systems. In the three-dimensional case this 
eigenvalue is two-fold degenerate.

Also in the three-sphere system the semiclassical zeta function can be
constructed from the corresponding expression for the two-dimensional
three-disk system. As all periodic orbits lie in one plane, the $4
\times 4$ monodromy matrix decomposes into blocks of $2 \times 2$
matrices, where the off-diagonal blocks vanish. One block describes
the motion in the plane and it is given by the monodromy matrix of the
three-disk system. The other block describes the motion perpendicular
to the plane and can be constructed from the same section lengths
$l_i$, local curvature radii $\rho_i$ and scattering angles $\theta_i$
as in the three-disk system.  The only difference is that the matrix
elements of the $2\times 2$ dimensional reflecting matrix of
Ref.\cite{Pinball} change to~\cite{Primack_priv}
\be 
\left ( \begin{array}{cc} -1 & -2/[\rho_i \cos(\theta_i)] \\ 0 
      & -1 \end{array} \right ) \longrightarrow \left (
\begin{array}{cc} 1 & 2 \cos(\theta_i) /\rho_i \\ 0 & 1 \end{array}
\right ) \; , \ee whereas the translational matrices stay as
\[
 \left ( \begin{array}{cc} 1 & 0 \\
                           l_i  & 1              \end{array} \right ) \ .
\]
Note that in the 2-scatterer cases the angles $\theta_i$ are always zero.

In Fig.\,11 the quantum-mechanically and semiclassically
calculated resonances of the completely symmetric representation of
the two-sphere system are shown. The agreement between the
quantum-mechanical and semiclassical data in the leading Gutzwiller
band is very good except for the first few resonances located at small
real parts of the wave number $k$. In contrast, the suppressed quantum
resonances situated in the diffraction band cannot be described by
these semiclassical calculations. Note that the agreement of the
leading resonances calculated quantum-mechanically and semiclassically
is already very good at quite small real parts of $k$. For increasing
${\rm Re}\,k$ this agreement improves.

Also the quantum-mechanical and semiclassical resonances of the
completely symmetric representation of the three-sphere-system have
been calculated.  Again, we find a good agreement between the
quantum-mechanical and semiclassical data in the case of the leading
Gutzwiller band except for the first few low-lying resonances. As in
the two-sphere case the suppressed quantum resonances of the
diffraction band cannot be described by these semiclassical
calculations.

In the four-sphere-system, we have only determined the three
fundamental periodic orbits so far. Even in this case a comparison
between the zeros of the quantum determinant $\Det \mbf{M}$ calculated
via the cumulant expansion (\ref{Cumulant_entw}) which has been
truncated after the first cumulant shows an agreement between the
quantum and semiclassical leading resonances.

\section{Conclusions}

We presented a quantum mechanical and a semiclassical discussion of a
chaotic system in three dimensions.  As a model, we chose the
scattering of a point particle on $N$ hard spheres in three
dimensions.  We showed that certain methods which were developed in
the framework of two-dimensional systems can be extended to three
dimensions in a straightforward way. Within the framework of
stationary scattering theory and using Green's theorem an expression
for the determinant of the scattering matrix {\bf S} was derived. In
order to determine the scattering resonances it suffices to look for
the poles of the determinant of the {\bf S}-matrix.  The determinant
of the {\bf S}-matrix of the entire $N$-sphere scattering system
splits up into a coherent and an incoherent part. The coherent part is
given by the ratio of the determinants of the hermitian conjugate and
of the characteristic matrix {\bf M} itself, where {\bf M} contains
the information on the geometry of the entire $N$-scatterer
configuration. The incoherent part consists of the product of the $N$
determinants of the 1-scatterer {\bf S}-matrices, each expressed in a
coordinate system whose origin lies in the center of the individual
scatterer. The expression for $\det \mbf{S}^{(N)}$ respects the
unitarity of the scattering matrix and guarantees a unitary
semiclassical limit without any resummation techniques a la Berry and
Keating~\cite{berry_keats}.  The scattering resonances are given by
the zeros of the determinant of {\bf M}.  The poles of the incoherent
part of $\det \mbf{S}^{(N)}$ get canceled by the poles of $\Det
\mbf{M}$.  A proof was given that all the formal manipulations in our
derivations are allowed and that the final expression for $\det
\mbf{S}^{(N)}$ is well defined provided that the number of scatterers
is finite and that they do not touch nor overlap. Similar results have
been found in the description of two-dimensional $N$-disk scattering
systems~\cite{GaRiqm,unspaper}.

We conjecture the following direct link between the quantum-mechanical
and semiclassical descriptions of $N$-sphere scattering systems: The
semiclassical limit of $\Tr(\mbf{W}^n)$, $\mbf{W} \equiv \mbf{M} -
\mbf{1}$, is given by those terms of the curvature regulated
Gutzwiller-Voros zeta function which correspond to periodic orbits of
total topological length $n = n_pr$ ($r$ denotes the number of
repeats), each weighted with the topological length $n_p$ of the
underlying primitive periodic orbit, plus diffractional corrections.
As the determinant of $\mbf{M}$ is given by the cumulant expansion,
this means that the semiclassical limit of the $n$-th cumulant of
$\Det \mbf{M}$ is given by the $n$-th curvature order of the
Gutzwiller-Voros zeta function plus diffractional corrections. This
connection holds for all complex wave numbers.  It is direct, as it
does not rely on the subtraction of the (integrated) spectral
densities of two equally sized infinitely large {\em bounded}
reference systems -- one containing the $N$-sphere scatterer and the
other not~\cite{Pinball,BaBl}.  Again this is analogous to the results
found in $N$-disk scattering systems~\cite{unspaper}.

Qualitatively, the distribution of the quantum-mechanically calculated
resonances of $N$-sphere systems is similar to the known results of
$N$-disk systems~\cite{GaRiqm,unspaper}. Comparing the computed
resonances of analogous $N$-sphere and $N$-disk systems, which have all
the geometric periodic orbits in common, we see that the leading
Gutzwiller resonances have the same real part but the whole resonance
band of the three-dimensional system is shifted to smaller imaginary
parts. This is due to the increased instability of the geometrical
periodic orbits in the three-dimensional case. In the case of the
suppressed resonances located in diffraction bands there is no such
correspondence.  The comparison of quantum-mechanically and
semiclassically calculated $N$-sphere resonances shows qualitatively the
same behaviour as in the two-dimensional $N$-disk systems.

The semiclassical investigation of $N$-sphere scattering systems is
still a sparsely studied field. The description of diffractional
corrections could be the subject of future projects.  In this context
an investigation of $N$-sphere systems in which the sphere separation is
much bigger than the radii of the scatterers would be of interest,
too, because diffraction effects should become more important.  Also
the opposite case with configurations of almost touching scatterers
provides an interesting system in which bound states might develop.
Another important point is a further investigation of the here
conjectured link between the quantum-mechanical and semiclassical
descriptions.

\section*{Acknowledgment}

We would like to thank Friedrich Beck and Predrag Cvitanovi\'c for
fruitful discussions and helpful advice.  We are grateful to Harel
Primack for sending us the structure of the three-dimensional
$N$-sphere Jacobians.  A.W. acknowledges the warm hospitality of the
Center of Chaos and Turbulence Studies at the Niels Bohr Institute in
Copenhagen, Denmark, where part of this work was done.  T.G. thanks
the Danish Research Council for financial support.

\appendix

\section{Details of the Calculation of the Scattering Matrix}

We distinguish between the two possible cases
defined in Sec.\,2.2.

{\bf First case:} $\vec{r}{\,}' \equiv \vec{r}_j \in \mbox{boundary 
of the $j^{\rm \, th}$
scatterer}$ \\ \nopagebreak
\begin{samepage}
Green's theorem can be written in the form 
\be    \label{Green_a}
     0 = I^j_\infty + \sum_{j'=1}^N I^j_{j'}  \; ,
\ee
where the integrals are given by
\bee    \label{int_def_a}
   I^j_\infty & = & \int_{\partial_\infty V} d\vec{a} \cdot
    [\psi_{l m}^k(\vec{a}) \vec{\nabla}
    G(\vec{a},\vec{r}_j)
       - G(\vec{a},\vec{r}_j) \vec{\nabla} \psi_{l m}^k(\vec{a})]  \\
    I^j_{j'}        & = & - \int_{\partial_{j'} V} d\vec{a} \cdot
         G(\vec{a},\vec{r}_j) \vec{\nabla} \psi_{l m}^k(\vec{a}) \; .
\eee
%
The integrals defined in Eqs.~(\ref{int_def_a}) can be worked
out in a straightforward calculation, we arrive at
\bee  
   I^j_\infty & = & -{\sqrt{4\pi}}^5 \sum_{l_1,l_2=0}^\infty
                  \sum_{\tilde{m},m_2=-l_2}^{l_2}
     {(-1)}^m i^{l_1+l_2}\sqrt{(2l+1)(2l_1+1)(2l_2+1)} \nn \\   
 && \quad \mbox{} \times
                  j_{l_1}(ks_j)j_{l_2}(ka_j)
      \dreij{l_1}{l_2}{l}{0}{0}{0}\dreij{l_1}{l_2}{l}{m-m_2}{m_2}{-m} 
             \nn \\
 && \quad \mbox{} \times
       Y_{l_1,m-m_2}(\hat{s}_j)Y_{l_2,\tilde{m}}(\hat{a}_j^{(j)})
           D^{l_2}_{\tilde{m}m_2}(gl,j)  \; ,  
 \label{I-j-infty} \\
  \label{I_j_j}
     I_j^j &=& 4\pi ik \sum_{l'=0}^\infty \sum_{m'=-l'}^{l'}
     a_j^2 A^j_{lm,l'm'}
       \, j_{l'}(ka_j) \, h_{l'}^{(1)}(ka_j) \, Y_{l'm'}(\hat{a}^{(j)}_j)
\; , \\
  I^j_{j'} & = & {\sqrt{4\pi}}^3ika_{j'}^2 \sum_{l',l_1,l_2=0}^\infty
          \sum_{m'=-l'}^{l'}
   \sum_{m_2,\tilde{m}=-l_2}^{l_2}{(-1)}^{m'}i^{l_1+l_2-l'} \,
   \mbf{A}^{j'}_{lm,l'm'} 
\nn \\ 
 && \quad \mbox{} \times
   \sqrt{(2l'+1)(2l_1+1)(2l_2+1)}
   \dreij{l_1}{l_2}{l'}{0}{0}{0}\dreij{l_1}{l_2}{l'}{m'-m_2}{m_2}{-m'}
           \nn  \\ 
 && \quad \mbox{} \times
  j_{l'}(ka_{j'}) \, j_{l_2}(ka_j) \,
        h^{(1)}_{l_1}(kR_{j'j})
   Y_{l_1,m'-m_2}(\hat{R}^{(j')}_{j'j}) \, Y_{l_2\tilde{m}}(\hat{a}_j^{(j)})
   \, D^{l_2}_{\tilde{m}m_2}(j',j)   \; .  \label{I_j_jpr}
\eee
In the derivation we made use of the addition theorems for Bessel
functions,
\bee  
    i^l j_l(kr) Y_{lm}(\hat{r}) & = & \sqrt{4\pi} \sum_{l_1,l_2=0}^\infty
      \sum_{m_1=-l_1}^{l_1} {(-1)}^m i^{l_1+l_2} \nn \\ 
 && \ \mbox{} \times 
      \sqrt{(2l+1)(2l_1+1)(2l_2+1)}j_{l_1}(ks) \, j_{l_2}(ka) 
       \nn \\ 
 && \ \mbox{} \times
      \dreij{l_1}{l_2}{l}{0}{0}{0}\dreij{l_1}{l_2}{l}{m_1}{m-m_1}{-m}
      Y_{l_1m_1}(\hat{s})Y_{l_2,m-m_1}(\hat{a}),  \label{bessel_add} 
         \\
    \vec{r} & = & \vec{s} + \vec{a} \; , \nn
\eee
\end{samepage} 
\newpage
and Hankel functions \cite{Abramovitz},
\bee    \label{hankel_add}
    i^l h^{(1)}_l(kr) Y_{lm}(\hat{r}) & = & 
      \sqrt{4\pi} \sum_{l_1,l_2=0}^\infty
      \sum_{m_1=-l_1}^{l_1} {(-1)}^m i^{l_1+l_2} \nn \\ 
 && \  \mbox{} \times
      \sqrt{(2l+1)(2l_1+1)(2l_2+1)}h^{(1)}_{l_1}(ks)j_{l_2}(ka) 
        \nn \\
 && \  \mbox{} \times
      \dreij{l_1}{l_2}{l}{0}{0}{0}\dreij{l_1}{l_2}{l}{m_1}{m-m_1}{-m}
      Y_{l_1m_1}(\hat{s})Y_{l_2,m-m_1}(\hat{a}) ,               \\
    \vec{r} & = & \vec{s}+\vec{a} \quad , \quad s>a \; , \nn
\eee
which can be easily proven using $e^{i \vec{k} \cdot \vec{r}} = 
e^{i \vec{k} \cdot \vec{s}} e^{i \vec{k} \cdot \vec{a}}$ and the known
properties of Bessel and Hankel functions~\cite{Abramovitz}.
We use the definition of Ref.\,\cite{Landolt} for the $3j$-symbols.

For large distances from the scatterer configuration we use the
following asymptotic expressions for Hankel-functions~\cite{Abramovitz} ($kr
\to \infty$),
\be   \label{hankel_asymp}  
        h_l^{(2)}(kr)  \sim  \frac{1}{kr} e^{-i(kr - \frac{l+1}{2} \pi)}
             \; , \quad
        h_l^{(1)}(kr)  \sim  \frac{1}{kr} e^{+i(kr - \frac{l+1}{2} \pi)}
             \; .
\ee

In changing coordinate systems we adopt the definitions of Rose~\cite{Rose}
for the Euler angles and the irreducible representations of the rotational
group, e.g., 
$$
 D^j_{m'm}(\alpha \beta \gamma) =
     \bra{jm'}e^{-i \alpha J_z}e^{-i \beta J_y}e^{-i \gamma J_z}\ket{jm} \; ,
$$
 and
$D^l_{m'm}(gl,j)$ denotes the corresponding quantity where the Euler angles
describe the rotation of the axes of the $(j)$-system in those of the global
system. For the spherical harmonics $Y_{lm}$ we adopt the usual definition.

In particular, in the calculation of $I_\infty^j$ the integration is performed
in the global
coordinate system. The Green's function are evaluated under 
the asymptotic expression
(\ref{hankel_asymp}) and finally the addition theorem
(\ref{bessel_add}) is applied and the coordinate system is changed. This leads
to Eq.\,(\ref{I-j-infty}).
In the calculation of $I_j^j$ the integration is performed in the
$(j)$-system. One easily obtains Eq.\,(\ref{I_j_j}).
In the calculation of $I^j_{j'}$, $j \neq j'$, we first perform the angular 
integration 
in the $(j')$-system. Then the addition theorem (\ref{hankel_add}) is used and
finally the coordinate system is changed. We get Eq.\,(\ref{I_j_jpr}).

The expressions (\ref{I-j-infty}) to (\ref{I_j_jpr}) are already written 
in a form $\sum_{lm}
\; \cdots \; Y_{lm}(\hat{a}_j^{(j)})$. It is now easy to see that
Eq. (\ref{Green_a}), written as $- I^j_\infty  =  \sum_{j'=1}^N I^j_{j'}$,
represents an equality between functions defined on the surfaces of the
$j$-th scatterer and that it becomes 
\bee  \label{M_entstehung}
   \sum_{l_2,m_2} \tilde{\mbf{C}}^j_{lm,l_2m_2} Y_{l_2m_2}(\hat{a}^{(j)}_j) 
      & = &
   \sum_{l_2,m_2} \sum_{j',l',m'} \mbf{A}^{j'}_{lm,l'm'}
     \tilde{\mbf{M}}^{j'j}_{l'm',l_2m_2}  Y_{l_2m_2}(\hat{a}^{(j)}_j) \; .
\eee

We have obtained an equality between the coefficients appearing in the
last equation up to a transformation of the form $\mbox{\bf C'} \equiv
\mbox{\bf CE} = \mbox{\bf AME} \equiv \mbox{\bf AM'}$.  Because of the
reasons already discussed in Secs.\,2 and~4 we write the
characteristic matrix as $\mbf{M} = \mbf{1} + \mbf{W}$ where the
diagonal entries of {\bf W} vanish.  With these definitions we obtain
Eqs.\,(\ref{CAM})--(\ref{M_N_ball_new}).

{\bf Second case:} $\vec{r}{\,}' \in V$, $r'$ large 
\\ \nopagebreak
According to (\ref{Green}) we obtain 
\be    \label{green_b_b}
    \psi^k_{lm}(\vec{r}{\,}'\,) = -\frac{1}{4\pi} \left(  
   I_\infty^{\vec{r}{\,}'} +
       \sum_{j=1}^N I_j^{\vec{r}{\,}'} \right) 
\ee
where the integrals are given by
\bee   \label{I_def_b}
   I^{\vec{r}{\,}'}_\infty & = & \int_{\partial_\infty V} d\vec{a} \cdot
     (\psi^k_{lm}(\vec{a}) \vec{\nabla} G(\vec{a},\vec{r}{\,}'\,) -
     G(\vec{a},\vec{r}{\,}'\,) \vec{\nabla} \psi^k_{lm}(\vec{a}))  \\
   I^{\vec{r}{\,}'}_j & = & - \int_{\partial_j V} d\vec{a} \cdot
      G(\vec{a},\vec{r}{\,}'\,) \vec{\nabla} \psi^k_{lm}(\vec{a})
\eee
which yields Eq.~(\ref{green_b_b}) and thus determines the components 
of the wave function. The evaluation of the
integrals yields
\bee    \label{I_r_infty}
    I_{\infty}^{\vec{r}{\,}'} &=& -16{\pi}^2 i^l j_l(kr') Y_{lm}(\hat{r}')\\
   I_j^{\vec{r}{\,}'} & = & (4\pi)^{\frac{3}{2}} ika_j^2 \sum_{l'=0}^\infty
    \sum_{m'=-l'}^{l'} \sum_{l_1,l_2=0}^\infty \sum_{m_1=-l_1}^{l_1}
    (-1)^{m'+l_2} i^{l_1+l_2-l'} \nn \\ 
  && \quad \mbox{} \times
   \sqrt{(2l'+1)(2l_1+1)(2l_2+1)} h^{(1)}_{l_1}(kr) j_{l_2}(ks_j) j_{l'}(ka_j)
     \dreij{l_1}{l_2}{l'}{0}{0}{0}   \nn  \\ 
   && \quad \mbox{} \times
   \dreij{l_1}{l_2}{l'}{m_1}{m'-m_1}{-m'} Y_{l_2,m'-m_1}(\hat{s}_j)
           Y_{l_1m_1}(\hat{r}') \tilde{\mbf{A}}^j_{lm,l'm'} \; . 
 \label{I_r_j}
\eee
The calculation of $I_\infty^{\vec{r}}$  can be carried out
in an analogous way as the determination of $I_\infty^j$.
In the calculation of $I_j^{\vec{r}}$ the integration is performed in the
$(j)$-system, then the addition theorem (\ref{hankel_add}) is used. This leads
to Eq.\,(\ref{I_r_j}).

One could insert the last results in Eq. (\ref{green_b_b}) and drop the
condition $kr' \gg 1$. This would yield an expression for the wave function
valid for all $r'$. The only restriction to $\vec{r}{\,}'$ is that the
addition theorem~(\ref{hankel_add}) of Hankel-functions has to hold. 
In order to
determine the {\bf S}-matrix it suffices to use $\vec{r}{\,}'$ far away from
the scatterers, such that Eq. (\ref{Def_S}) becomes valid. Hence, inserting
(\ref{green_b_b}) in (\ref{Def_S}) we obtain
\be    \label{S_1}
   \mbf{S}_{lm,l'm'} = \delta_{ll'} \delta_{mm'} -
                  i\sum_{j'',l'',m''}
       \tilde{\mbf{A}}^{j''}_{lm,l''m''} 
        \tilde{\mbf{D}}^{j''}_{l''m'',l'm'} \; .
\ee
We use $\tilde{\mbf{A}}^j_{lm,l'm'} = \sum_{\tilde{m}=-l'}^{l'}  
\mbf{A}^j_{lm,l'\tilde{m}}
               D^{l'}_{m'\tilde{m}}(j,gl)$  to get 
this result compatible with
the results obtained in the first case. By this operation the $N$ local
coordinate systems appear and we obtain Eqs.~(\ref{S_2}) and (\ref{D_new}).


\section{Symmetry Considerations}

In order to simplify the determination of scattering resonances, given
as the zeros of the determinant of the characteristic matrix {\bf M},
the symmetry of the scattering configuration can be used to
block-diagonalize {\bf M}.  (These operations are allowed as $\mbf{M}-
\mbf{1}$ has been proven trace-class and as $\Det \mbf M$ exists.)  A
similar approach was used in Ref.\,\cite{Cv_Eck} in order to perform
symmetry reductions in $N$-disk scattering systems in a semiclassical
description.

The infinite square matrix {\bf M} can be interpreted as a linear application
acting on the square integrable functions defined on the surfaces of the $N$
scatterers. In Sec.\,2 we have seen that {\bf M} contains the complete
information on the geometry of the entire scattering system, so we have
\be   \label{M_komm_rho}
   \left[ \mbf{M},\rho (g) \right]  =  0 \; , \qquad  \forall g \in G \; .
\ee

Here $\rho$ denotes the linear representation of the symmetry group $G$ acting
in the same space as {\bf M}. As the entire scattering configuration can be
constructed by successive applications of the symmetry transformations of $G$
acting on an appropriate fundamental domain, the representation $\rho$ is
decomposable,
\be    \label{rho_zerl}
   \rho = R \otimes \tilde{\rho} \; .
\ee

The representation $R$ acts on the $|G|$ ($=$ number of elements of $G$)
copies of the fundamental domain and $\tilde{\rho}$ acts on an appropriate
basis of the square integrable functions defined in the fundamental domain.
Now it is easy to see that $R$ is the regular representation~\cite{Hammermesh}
of $G$. If the symmetry group $G$ is finite, the regular representation is, as
is well known, reducible,
\be  \label{reg_Darst_zerl}
    R = \bigoplus_c d_c D_c \; .
\ee
Here the index $c$ runs over all conjugacy classes and $D_c$ is the $c$-th
irreducible representation of dimension $d_c$. Therefore, if we are dealing
with scattering systems that possess only discrete symmetries (this suffices
to ensure $|G| < \infty$ as we are dealing with finite, $N < \infty$,
systems) we obtain   
\be     \label{detM_zerl}
  \Det \mbox{\bf M} = \prod_c (\Det \tilde{\mbox{\bf M}}_{D_c})^{d_c}  \; .
\ee

The last result is inserted in Eq.\,(\ref{det_S}) and formula
(\ref{det_S_zerl}) is obtained.

\newpage

\newpage 
\pagestyle{empty}

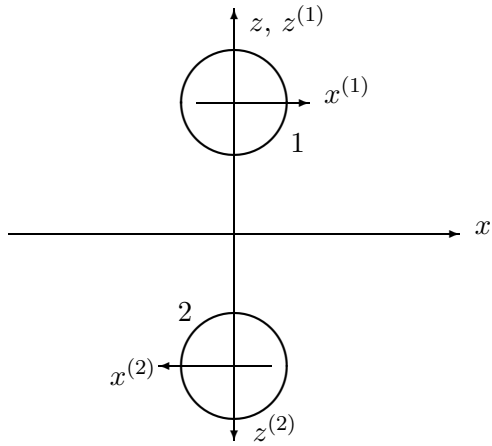
\begin{figure}
 \begin{center}
  \setlength{\unitlength}{0.5cm}
     \begin{picture}(20,12)

\put(4.0,6.0){\vector(1,0){12}}  \put(16.4,6.0){$x$}
\put(10.0,2.0){\vector(0,1){10}}  \put(10.4,11.4){$z$, $z^{(1)}$}

\thicklines  \put(10.0,9.5){\circle{3.0}} \thinlines
  \put(11.5,8.2){1}
  \put(9.0,9.5){\vector(1,0){3}}  \put(12.4,9.5){$x^{(1)}$}

\thicklines  \put(10.0,2.5){\circle{3.0}}  \thinlines
  \put(8.5,3.7){2}
  \put(11.0,2.5){\vector(-1,0){3}}  \put(6.7,2.0){$x^{(2)}$}
  \put(10.0,2.5){\vector(0,-1){2}}  \put(10.5,0.5){$z^{(2)}$}

     \end{picture}
\caption[two-sphere-configuration]{The two-sphere-configuration 
consisting of two
equal spheres. The global and local coordinate systems used are also
shown.}\label{Abb7} 

  \end{center}
\end{figure}
  
\newpage

\begin{samepage}
\begin{figure}[b]
  \epsfig{file=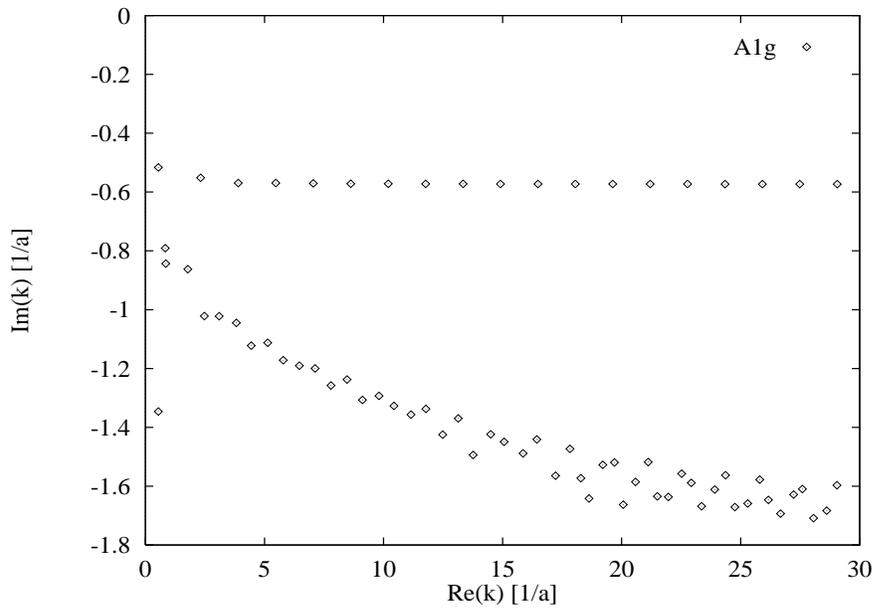,width=12cm,height=8cm,angle=-90}
  \caption[two-sphere, $A_{1g}$]{
Quantum-mechanically calculated $A_{1g}$-resonances of
the two-sphere system
in the complex $k$-plane.
The center-to-center seperation is $R=6a$ and the wave number is measured in
units of the inverse sphere radius~$a$. }\label{Abb15a}

\end{figure}

\begin{figure}[b]
  \epsfig{file=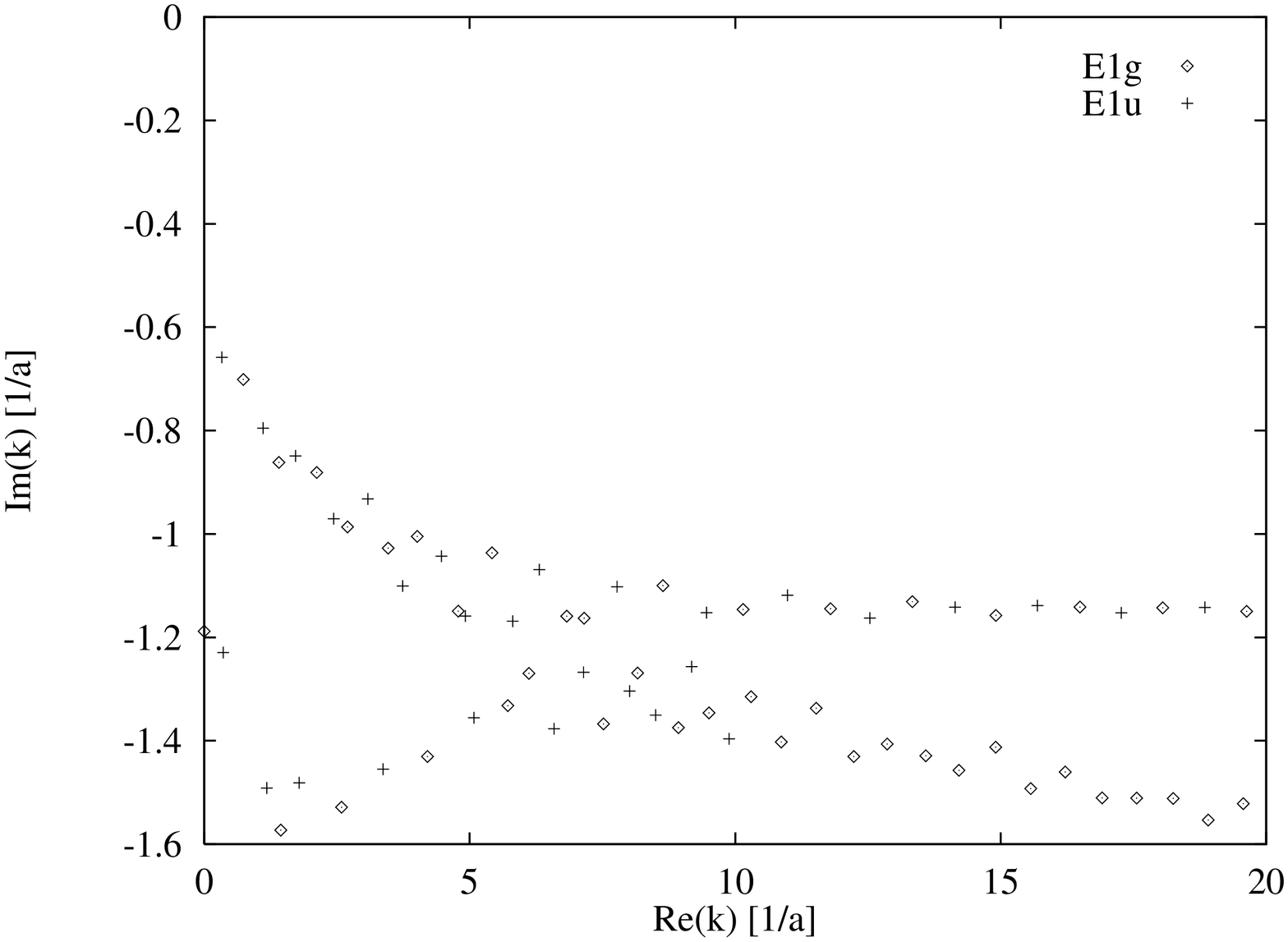,width=12cm,height=8cm,angle=-90}
  \caption[two-sphere, $E_{1g}$ und $E_{1u}$]{Quantum-mechanically calculated
$E_{1g}$- and $E_{1u}$-resonances of the two-sphere system
in the complex $k$-plane. }\label{Abb15b}

\end{figure}
\end{samepage}
                 
\thispagestyle{empty}

\begin{figure}[b]
  \epsfig{file=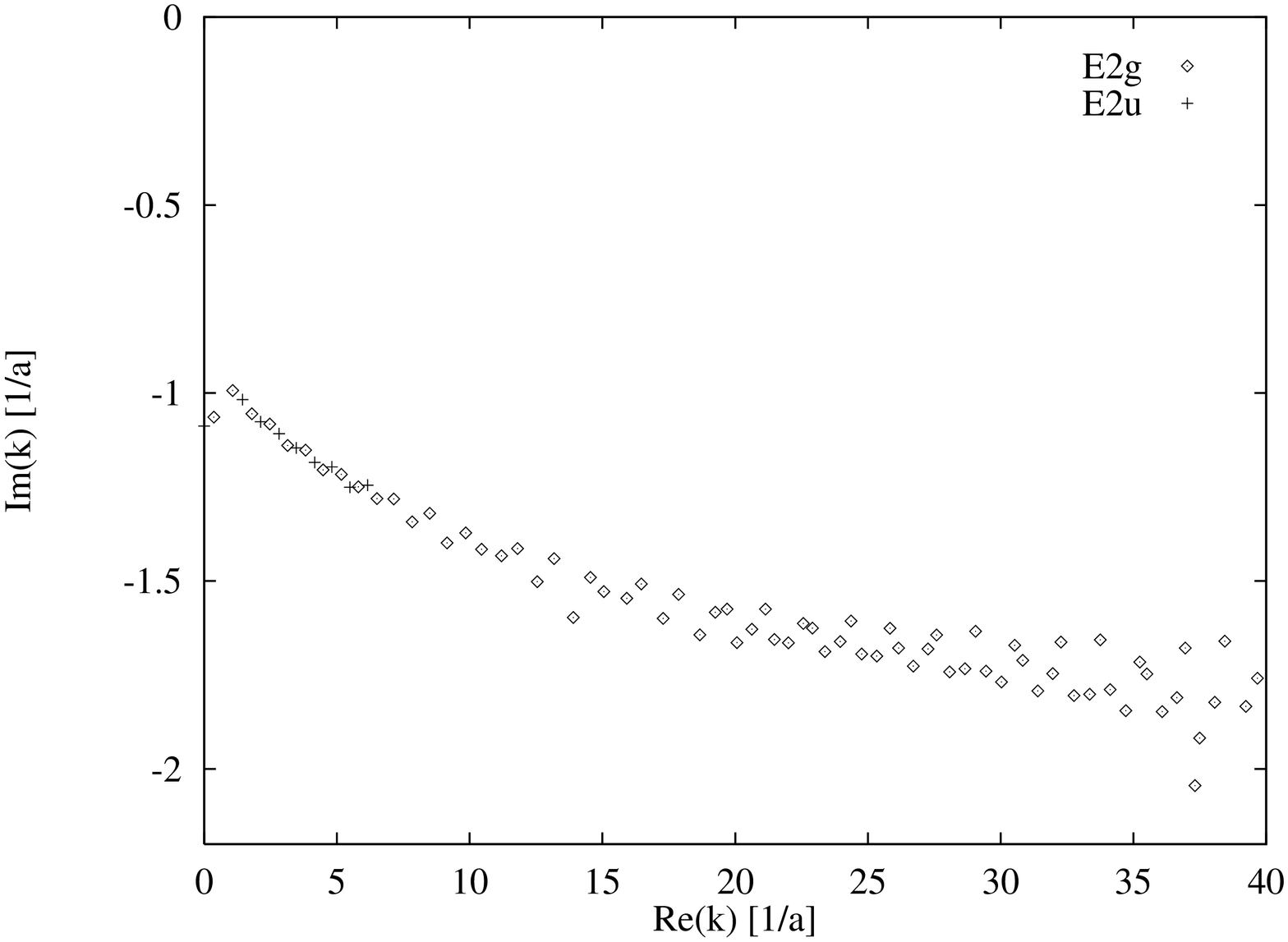,width=12cm,height=8cm,angle=-90}
  \caption[two-sphere, $E_{2g}$ und $E_{2u}$]{
Quantum-mechanically calculated $E_{2g}$- and $E_{2u}$-resonances
of the two-sphere system in the complex $k$-plane. }\label{Abb15c}

\end{figure}

\newpage 
\thispagestyle{empty}

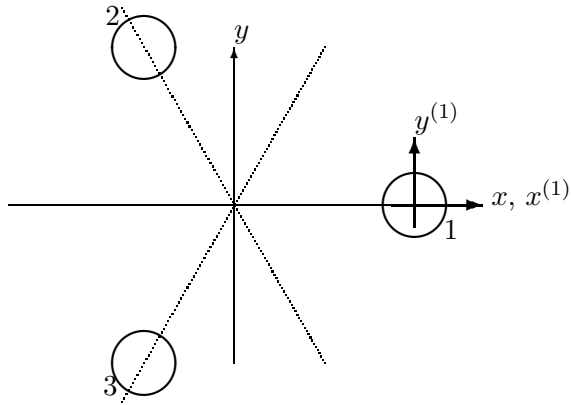
\begin{figure}
 \begin{center}
  \setlength{\unitlength}{0.3cm}
     \begin{picture}(24,16)
 \put(2.0,9.0){\vector(1,0){21}}
    \put(23.4,9.0){$x$, $x^{(1)}$}
 \put(12.0,2.0){\vector(0,1){14}}
    \put(12.0,16.4){$y$}

\thicklines
 \put(20.0,8.0){\vector(0,1){4}}
    \put(20.0,12.4){$y^{(1)}$}
    \put(19.0,9.0){\vector(1,0){4}}
\put(20.0,9.0){\circle{3.0}}  \put(21.3,7.5){1}

\put(8.0,16.0){\circle{3.0}}  \put(6.3,17.0){2}

\put(8.0,2.0){\circle{3.0}}   \put(6.2,0.6){3}
\thinlines

 \bezier{100}(4.0,9.0)(12.0,9.0)(22.0,9.0)

 \bezier{100}(7.0,17.75)(12.0,9.0)(16.0,2.0)
 \bezier{100}(7.0,0.25)(12.0,9.0)(16.0,16.0)

      \thinlines

    \end{picture}
\caption[three-sphere-configuration]{The three-sphere-configuration 
consisting of three equal
spheres. The global and the local coordinate system of the first sphere are
also shown. The remaining two local coordinate systems are obtained by
rotations through $\frac{2 \pi}{3}$ (sphere two) and $\frac{4 \pi}{3}$ 
(sphere three)
around the origin of
the global system. All $z$-axes are perpendicular to the figure
plane.}\label{Abb8}

  \end{center}
\end{figure}

\newpage 
\thispagestyle{empty}

\begin{figure}[b]
  \epsfig{file=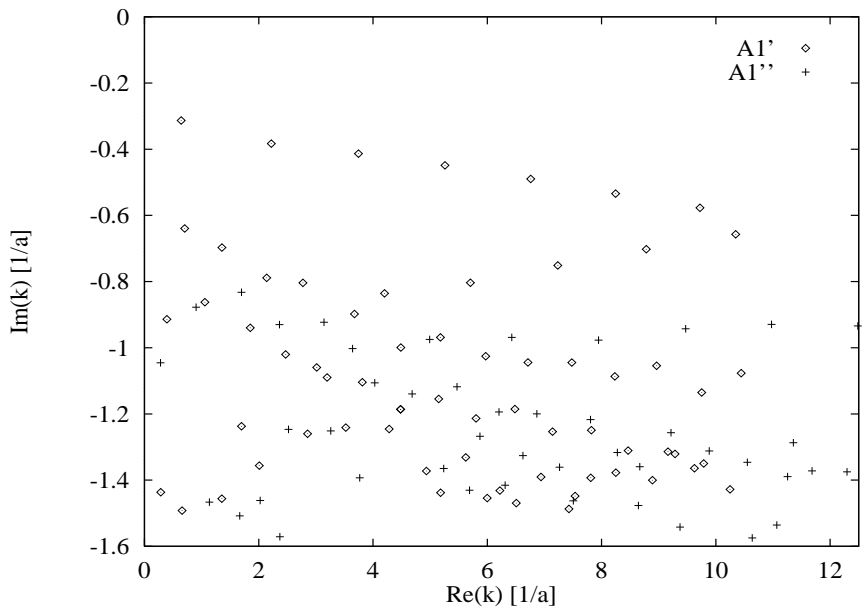,width=12cm,height=8cm,angle=-90}
  \caption[three-sphere, $A_1{}'$ und $A_1{}''$]{
Quantum-mechanically calculated $A_1{}'$- and $A_1{}''$-resonances of
the three-sphere-system in the complex $k$-plane. }\label{Abb17a}
\end{figure}

\newpage \thispagestyle{empty}

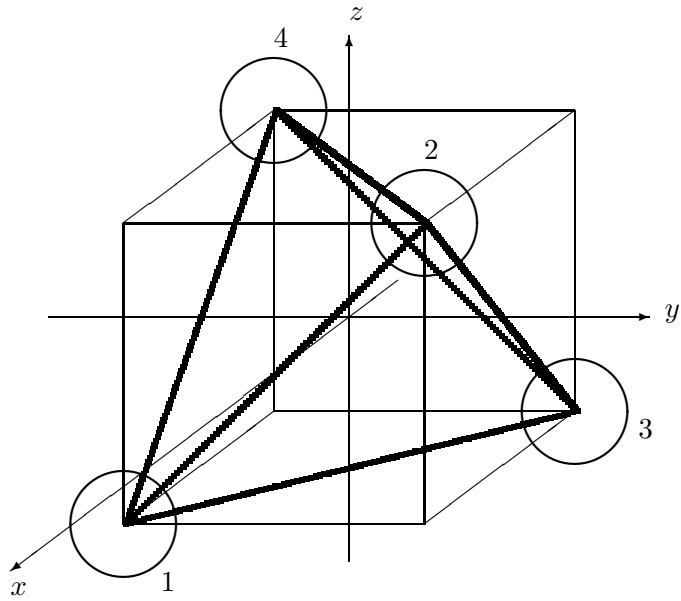
\begin{figure}[t]
 \begin{center}
  \setlength{\unitlength}{0.5cm}
     \begin{picture}(24,16)

\put(6.0,3.0){\line(1,0){8}}
\put(6.0,3.0){\line(0,1){8}}
\put(6.0,3.0){\line(4,3){4}}

\put(6.0,11.0){\line(1,0){8}}
\put(6.0,11.0){\line(4,3){4}}

\put(10.0,14.0){\line(1,0){8}}
\put(10.0,14.0){\line(0,-1){8}}

\put(14.0,11.0){\line(0,-1){8}}
\put(14.0,11.0){\line(4,3){4}}

\put(18.0,14.0){\line(0,-1){8}}

\put(14.0,3.0){\line(4,3){4}}

\put(10.0,6.0){\line(1,0){8}}

\thicklines
  \put(6.0,3.0){\circle{3.0}}     \put(7.0,1.2){1}

  \put(10.0,14.0){\circle{3.0}}   \put(10.0,15.7){4}

  \put(14.0,11.0){\circle{3.0}}   \put(14.0,12.7){2}

  \put(18.0,6.0){\circle{3.0}}    \put(19.7,5.3){3}

\linethickness{0.7mm}

 \bezier{100}(6.0,3.0)(12.0,4.5)(18.0,6.0)

 \bezier{100}(6.0,3.0)(8.0,8.5)(10.0,14.0)

 \bezier{100}(6.0,3.0)(10.0,7.0)(14.0,11.0)

 \bezier{100}(10.0,14.0)(12.0,12.5)(14.0,11.0)

 \bezier{100}(10.0,14.0)(14.0,10.0)(18.0,6.0)

 \bezier{100}(14.0,11.0)(16.0,8.5)(18.0,6.0)

\thinlines

\put(4.0,8.5){\vector(1,0){16}}   \put(20.4,8.5){$y$}

\put(12.0,2.0){\vector(0,1){14}}   \put(12.0,16.4){$z$}

\put(12.0,8.5){\vector(-4,-3){9}}   \put(3.0,1.1){$x$}
\put(12.0,8.5){\line(4,3){1.3}}

    \end{picture}

\caption[four-sphere-configuration] {The four-sphere-configuration
consists of four equal spheres at the corners of a regular tetrahedron
(thick lines). In order to have a better visualization of the
symmetries the tetrahedron is placed in a cube in whose center lies
the origin of the global coordinate system ($x,y,z$).}\label{Abb9}

 \end{center}
\end{figure}

\newpage 
\thispagestyle{empty}

\begin{samepage}

\begin{figure}[b]
  \epsfig{file=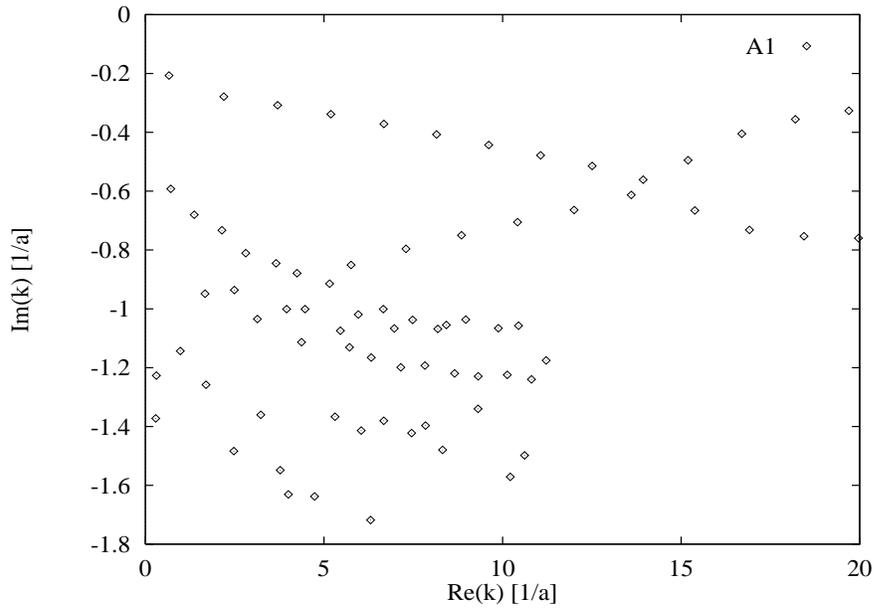,width=12cm,height=8cm,angle=-90}
  \caption[four-sphere, ${A_1}$]{Quantum-mechanically calculated
$A_1$-resonances of the four-sphere system in the complex $k$-plane.
For numerical reasons only the first subleading resonances have
been calculated. }\label{Abb19}

\end{figure}

\begin{figure}[b]
  \epsfig{file=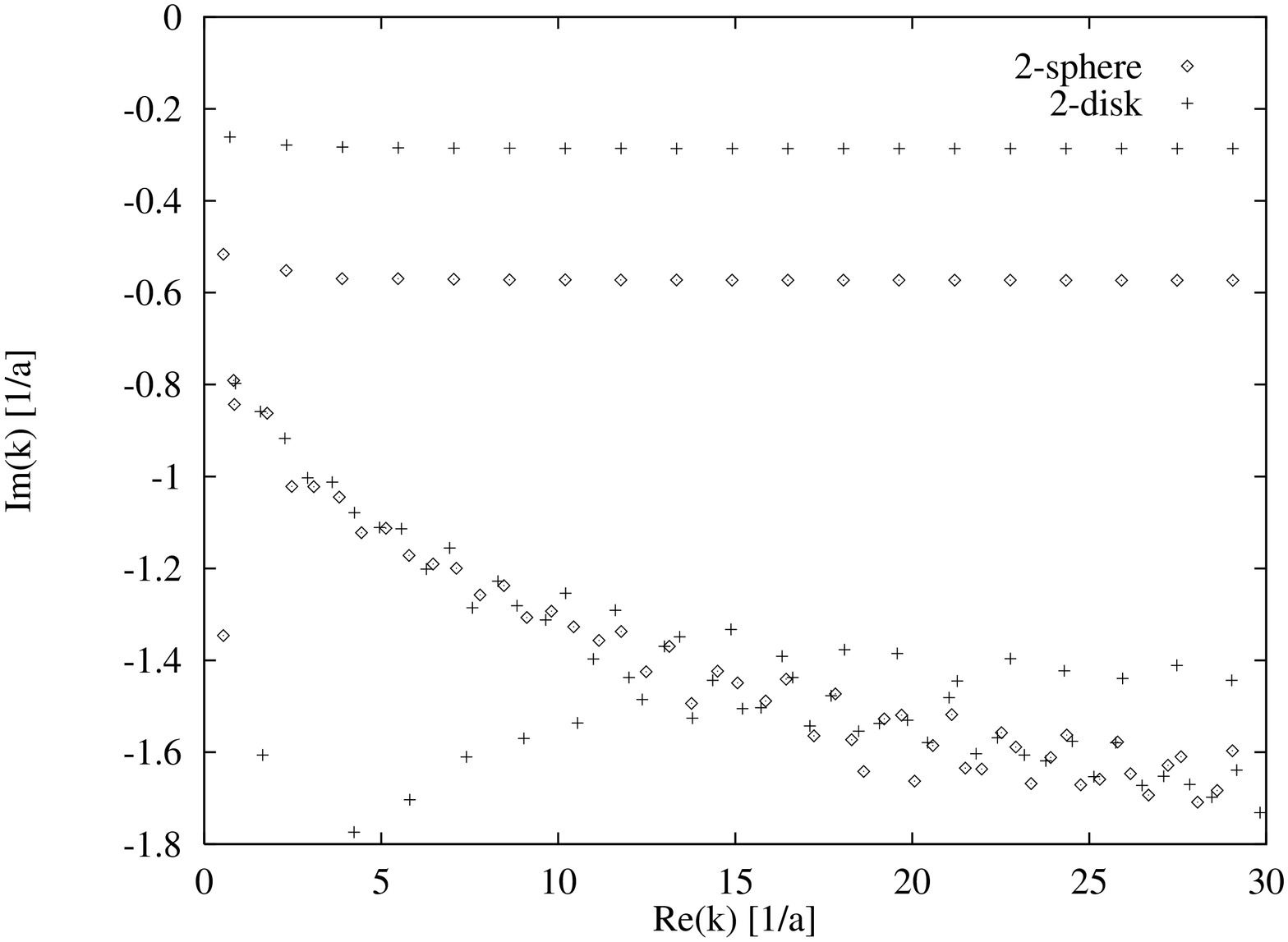,width=12cm,height=8cm,angle=-90}
  \caption[two-sphere versus two-disk]{
Comparison of the quantum-mechanically calculated resonances in the complex
$k$-plane of the totally
symmetric representations of the two-sphere and two-disk systems.
          }\label{Abb24}
\end{figure}
\end{samepage}

\begin{samepage}
\begin{figure}[b]
  \epsfig{file=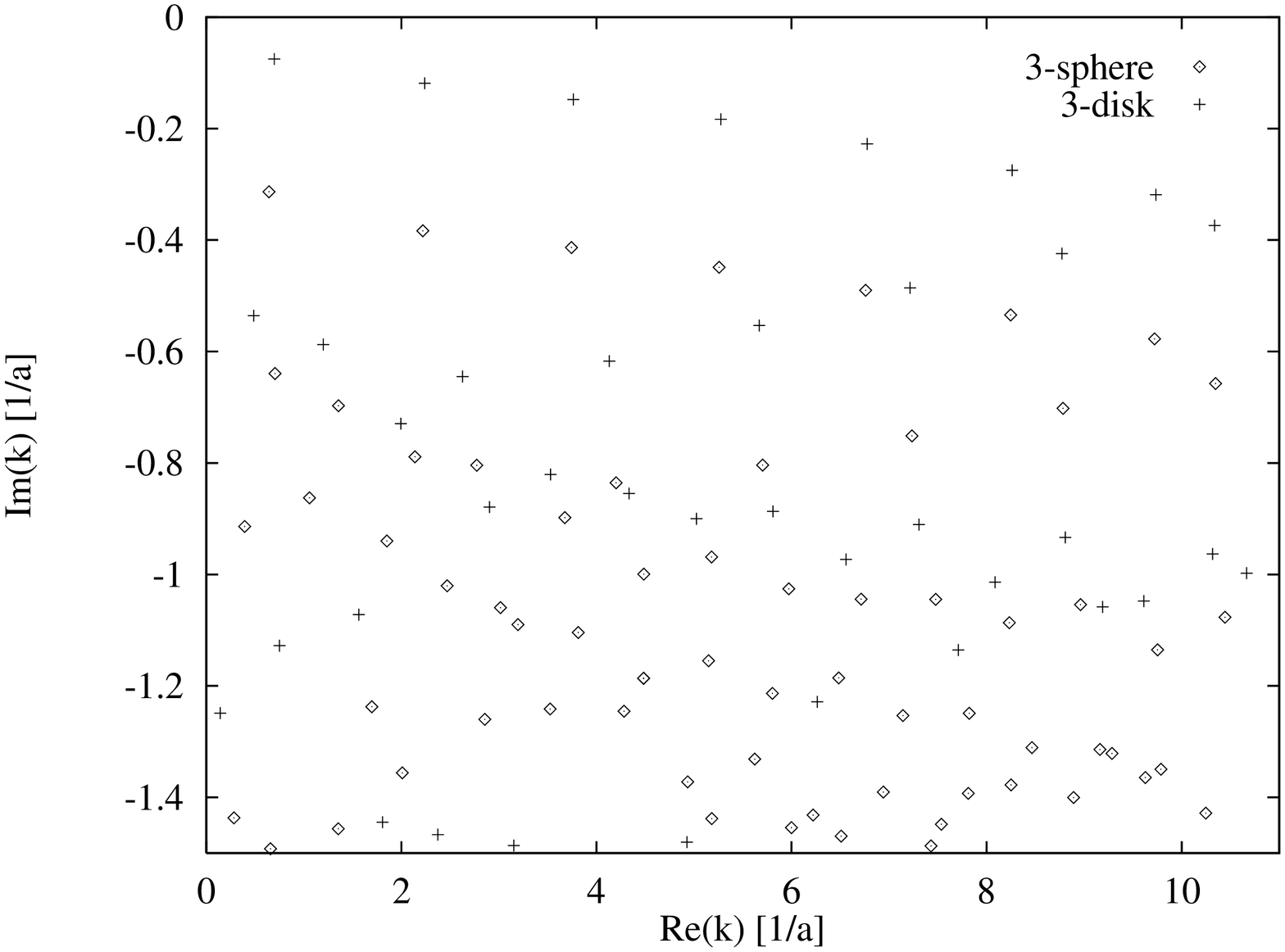,width=12cm,height=8cm,angle=-90}
  \caption[three-sphere versus three-disk]{
Comparison of the quantum-mechanically calculated resonances in the complex
$k$-plane of the totally
symmetric representations of the three-sphere and three-disk systems.
          }\label{Abb26}
\end{figure}

\pagebreak

\pagebreak

\begin{figure}[b]
  \epsfig{file=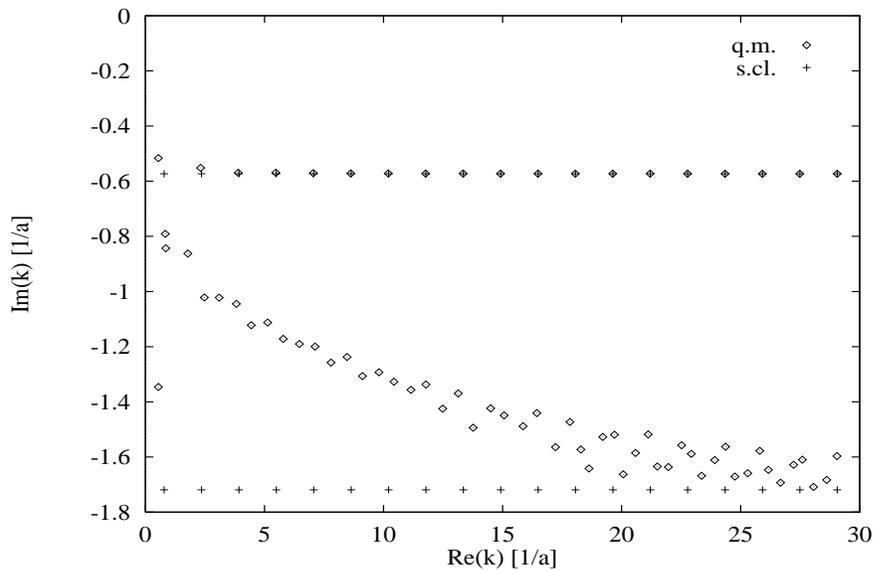,width=12cm,height=7.5cm,angle=-90}
  \caption[Vergleich q.m. --- s.kl., two-sphere]{
Comparison of quantum-mechanically~({\rm q.m.}) and semiclassically~({\rm
s.cl.}) calculated resonances of the completely symmetric representation of
the two-sphere system in the complex $k$-plane.
         }\label{Abb30}
\end{figure}
\end{samepage}

\end{document}